\documentclass[11pt]{article}
\usepackage{graphicx}
\usepackage{endnotes}
\usepackage{hyperref}
\usepackage{natbib}
\usepackage{amsmath,amsfonts,amsthm,bm}

\let\footnote=\endnote

\newcommand\hbindex[1]{#1}

\def\R{\ifmmode{I\hskip -3pt R}
           \else{\hbox{$I\hskip -3pt R$}}\fi}

\def\N{\ifmmode{I\hskip -3pt N}
           \else{\hbox{$I\hskip -3pt N$}}\fi}

\title{{\bf Probability Models in Statistical Data Analysis: Uses, Interpretations, Frequentism-As-model}}
\author{Christian Hennig,
 Dipartimento di Scienze Statistiche ``Paolo Fortunati'',\\ Universita di Bologna,\\
 Via delle Belle Arti, 41, 40126 Bologna\\
 christian.hennig@unibo.it}

\begin{document}

\maketitle
{\bf Abstract:}~\\
{\bf Note:} Published now as a chapter in ``Handbook of the History and Philosophy of Mathematical Practice'' (Springer Nature, editor B. Sriraman,\\ 
\verb|https://doi.org/10.1007/978-3-030-19071-2_105-1|)
~\\~\\
The application of mathematical probability theory in statistics is
quite controversial. Controversies regard both the interpretation of probability, and approaches to statistical inference. After having given an 
overview of the main approaches, I will propose a re-interpretation of 
frequentist probability. 
Most statisticians are aware that probability models interpreted
in a frequentist manner are not really true in objective reality, but 
only idealisations. I argue that this is often ignored when actually
applying frequentist methods and interpreting the results, and that keeping
up the awareness for the essential difference between reality and models can 
lead to a more appropriate use  and interpretation of frequentist models and 
methods, called ``frequentism-as-model''. 
This is elaborated showing connections to existing work, appreciating
the special role of independently and identically distributed 
observations and subject matter knowledge, giving an 
account of how and under what conditions models that are not true can be 
useful, giving detailed interpretations of tests and confidence intervals, 
confronting their implicit compatibility logic with the inverse probability 
logic of Bayesian inference,
re-interpreting the role of model assumptions, appreciating robustness, and
the role of ``interpretative equivalence'' of models.
Epistemic probability shares the issue that 
its models are only idealisations, and an analogous 
``epistemic-probability-as-model'' can also be developed.

\section{Introduction}
Ever since the time of Bernoulli \citep{Bernoulli1713}, mathematical 
\hbindex{probability} has been used for statistical inference. \cite{Kolmogorov33} has given probability a mathematical axiomatic foundation that is the basis of the mathematical theory of probability to this day. 

Whereas mathematical probability theory is largely uncontroversial, its application to statistical inference and data analysis is and has been highly contentious throughout the history of probability. This concerns both the very meaning of probability (e.g., \cite{Venn1866,Ramsey31,Galavotti05}), and how it can be used for learning from data about the world (e.g., \cite{BarCox62,WSL19,Mayo22}, the latter even referring to the ``statistics wars''). 

Here I will present a view of some of the central concepts of this controversy that is based on the constructivist philosophy of mathematical modelling elaborated in \cite{Hennig10}, along with some practical consequences. The main idea is that mathematical models are tools for thinking about reality, and they are essentially different from reality. Asking whether a mathematical model is true in reality constitutes a category error; mathematical models have other jobs than being true, and need to be assessed based on how well they do these other jobs. 

Although many statisticians tend to agree with George Box's famous quote that 
``all models are wrong but some are useful'' \citep{Box79} when asked directly,
the statistical literature and statistics teaching are full of statements of
the kind that ``model assumptions have to be fulfilled'', and references to 
``true distributions'' and ``true parameters'' in reality. Many philosophical
contributions to the foundations of statistics even more explicitly demand 
models to be true, or to be ``believed''. From the perspective of mathematical
modelling taken here, such a mindset easily leads to over-interpretation of
the results of statistical analysis. It also opens up any position in the 
controversy about the foundations of statistics to criticism that could be 
avoided by a more modest and honest acknowledgement of the role and limitations
of the use of probability models. 

First I will give an overview over several key concepts, namely 
interpretations of probability, approaches to statistical inference, and
a constructivist philosophy of mathematical modelling. Then, 
the central part of this
chapter 
introduces ``frequentism-as-model'', a new re-interpretation of frequentist 
probability and inference separating appropriately models and reality, and
discusses many of its implications. 
An analogous version of epistemic probability, 
``epistemic-probability-as-model'', is sketched in less detail before
the chapter is concluded.

\section{Basic concepts}\label{sbasic}
The basic concepts for this chapter are different \hbindex{interpretations of 
probability}, different approaches to \hbindex{statistical inference}, and a 
\hbindex{constructivist 
philosophy} of the relation between \hbindex{mathematical models} and reality. Here
I give an overview of these. Note that in much literature the distinction
between interpretations of probability and approaches to inference is blurred.
Often, Bayesian reasoning is identified with a (subjective) epistemic 
interpretation of probability, but there are non-Bayesian approaches to
do inference with epistemic probabilities, and Bayesian reasoning can be 
combined with frequentist probabilities as well. 
\subsection{Interpretations of probability}\label{sinterprob}
Interpretations of probability connect probability as mathematical object with 
the real world by explaining what mathematical probability statements are 
supposed to mean. There are two major classes of interpretations of probability,
aleatory and epistemic.  
\subsubsection{Aleatory probability}\label{saleatory}
\hbindex{Aleatory probability}, sometimes referred to as ``\hbindex{chance}'', 
is a generic term for thinking of a 
probability as an observer-independent feature of what \cite{Hacking65} calls
``chance setup'' or a \hbindex{data generating process} in reality. 
The best known aleatory interpretation is \hbindex{frequentism}. In
frequentism, the chance setup is conceived as an experiment or a situation
that can be repeated in principle infinitely often, and the probability of an 
event, $P(A)$, say, is then thought of as the limiting relative frequency of the
occurrence of $A$ when the number of random \hbindex{repetitions} tends to infinity.
I will refer to chance setups as ``experiments'' in the following, in line
with the use of ``random experiments'' in the literature. This does not
necessarily imply controlled experimentation.
The \hbindex{axioms} of probability are motivated from mathematical 
properties of relative frequencies.

There have been various attempts to make this notion precise. The most famous 
of them are probably due to \cite{Venn1866} and \cite{vonMises39}. Both of them
acknowledge that frequentist probabilities are an idealisation. Their existence
cannot be assured based on finite sequences of observations; in fact any 
behaviour in any finite part of the 
sequence is compatible with any limit value from a mathematical point of view.
Requiring identical repetitions is an idealisation as well. This is 
bypassed by \cite{vonMises39} instead requiring the specification of an ideally
infinite ``\hbindex{Kollektiv}'' of experiments within which probability is defined rather than relying on ``identity''. \cite{Venn1866} and many later frequentists take
a hypothetical point of view; their probabilities refer to what would happen
if the experiment were to be repeated infinitely. \cite{Hajek09} gives 15
arguments against such a view, many of which have to do with subtleties of 
mathematical limits, but he also attacks the anti-empirical nature of the idea.
  
Another key difficulty is the definition of \hbindex{randomness}. Randomness
can be defined based on \hbindex{algorithmic complexity}, i.e., a sequence can count as 
random if it cannot be generated by certain finitely describable 
algorithms, see \cite{MartinLof66}, in which case he also could prove that
limiting relative frequencies exist. But also this cannot be checked given 
only a finite number of observations. \cite{Fine73} even showed, given suitable
definitions, that ``apparently random'' finite sequences will automatically 
look as if their relative frequencies ``apparently converge'', without any 
condition regarding what will really happen if that sequence is infinitely 
continued. Looking at a sequence of relative
frequencies of the occurrence of a set $A$ in a ``randomly enough''  
repeated experiment will give the impression that these 
relative frequencies stabilise. This is a major motivation to think that
frequentist probabilities might exist in the real world, but Fine's result shows
that the impression is an implication of what kind of sequences an observer
attributes randomness to, and does not imply anything about the real existence 
of a limiting frequency. 

An alternative aleatory concept is the interpretation of probabilities as 
``\hbindex{propensities}'', going back to \cite{Popper59b}. A propensity is conceived as
an objective feature of an experiment to have the tendency to produce certain
results. If indeed the experiment were repeated infinitely, the idea is that
it would produce relative frequencies according to its propensities, but there
is no necessity for this to happen for a propensity to be defined. In this way, 
the propensity interpretation of probability allows to speak of probabilities 
connected to single realisations of an experiment, and it does not have to be
concerned about issues with mathematical limits, although the propensity 
interpretation does not put the researcher in a better position to make 
empirical statements about the values or even the very existence of
probabilities compared to frequentism. 

There are several varieties of propensity interpretations, 
see \cite{Gillies00}. Gillies' own propensity approach basically postulates
that propensities behave according to the probability axioms, and in case
repetitions are observed, it can then be tested empirically whether outcomes
behave as they should behave given the postulated probability model. 
Such tests can of course not be 
very powerful if the number of repetitions is low.

The biggest problems with aleatory probability interpretations stem, in 
my view, from the idea that their use requires them to exist objectively
in reality. Treating mathematical models as fundamentally different from
reality, as I will outline below, an aleatory interpretation
of probability should refer to thinking of reality {\it as if} it behaved in 
this way, without postulating or believing that it really does, which cannot
be ensured, or even very convincingly argued, due to the lack of 
infinite random repetition in reality. I will call
this ``frequentism-as-model'' (in the following I will mostly refer to
``frequentism'', as this is in much more frequent use in the literature
than ``aleatory''). 
Obviously, there needs to be a justification
how models used in this way contribute to science. Many scientists think that
objectivity is a main reason to use aleatory probabilities, but they can be 
called ``objective'' only in the sense that they {\it refer} to 
observer-independent reality, not in the sense that their observer-independent
existence could be granted.

\subsubsection{Epistemic probability}   \label{sepistemic}
An \hbindex{epistemic probability} $P(A)$ refers to the belief or knowledge of an agent 
regarding the event $A$. Epistemic probabilities do not require any 
repeatability of experiments; they apply to single situations as well as to 
experiments that are part of a sequence. The probably simplest and 
historically original way to explain $P(A)$ in an epistemic way 
is as a \hbindex{betting} rate; the agent
would be willing to pay up to $P(A)$ for a gamble in which she receives 1 in
case that $A$ obtains and 0 otherwise. The \hbindex{axioms} of probability can be 
motivated from requirements of fair betting prices requiring ``\hbindex{coherence}'', 
i.e., bets on several events have to be offered in  such a way that an 
opponent cannot put together a system of bets so that the agent incurs sure 
loss. This was formalised by \cite{Ramsey31} and \cite{deFinetti37}.  

As for aleatory probability, there 
are various versions of the epistemic interpretation. \cite{Ramsey31} and 
\cite{deFinetti37} held that probabilities are subjective. An agent makes
subjective initial prior assignments of probability, and
probability theory 
imposes the coherence of these assignments, and how these
have to be coherently updated when new evidence occurs, namely following
Bayes's Theorem. Different agents can have different probabilities for the 
same event, even if they hold the same information. 

\hbindex{Subjective probabilities} may seem unsuitable for science if  
the aim of science is assumed to be producing 
objective statements that hold independently of the researcher.
\cite{Keynes21} interpreted probability as 
an extension of logic to uncertain propositions, the plausibility of which,
given the evidence, could be quantified as a probability. Such a probability
is epistemic, but there is no subjective agent; the probability is meant to
be determined by the evidence without further subjective input. Keynes 
constrained numerical probabilities to propositions concerning a finite number
of cases that are equiprobable, i.e., they have the same probability assigned
based on the ``\hbindex{Principle of Indifference}'' in case that for none of them there
are reasons to think of them as more likely than the others. Ramsey and de 
Finetti criticised this as far too restrictive, and there are further issues
with the Principle of Indifference, see \cite{Gillies00}. More recent
developments of logical and
objectivist epistemic interpretations of probabilities, e.g., \cite{Jaynes03},
are less restricted.

\hbindex{Bayesian updating} of \hbindex{prior} probability 
assessments (``priors'')
is central according to many
proponents of epistemic probability, which is therefore often identified 
with ``\hbindex{Bayesianism}''. At the time of \cite{Bayes1763}, there was no explicit
distinction between aleatory and epistemic probabilities. Bayes's original
paper is somewhat ambiguous on the matter, and his theorem holds as well
for aleatory probabilities. \cite{Williamson10}, despite 
calling his interpretation ``Bayesianism'', is an example for an epistemic
interpretation that does not rely on Bayesian updating. 

The major difference between objectivist and subjectivist epistemic 
interpretations is how the priors should be chosen.
The subjective approach may lead to potentially arbitrary choices that may
not be an appropriate basis for science, but in very many situations it is hard
if not impossible to impose a unique objective choice. Ideally, at some point 
there is no evidence. At this point, an ``\hbindex{informationless prior}'' is chosen, and
then any incoming evidence can be processed using Bayesian updating. However,
general agreement regarding how to choose
informationless priors does not exist, and there are
issues and paradoxes with them \citep{KaWa96}. Furthermore, much evidence 
comes in 
forms that an initially chosen model cannot easily incorporate. Objective 
probabilities might be seen as desirable, but the subjectivists can state that
they are an unachievable ideal in most situations. \cite{Berger06}, in a defence
of \hbindex{objective Bayesianism}, admits that full objectivity cannot be achieved,
particularly when invoking existing information that does not come in an
appropriately formalised way as for example knowledge about logically
impossible parameter values or observations. The 
term ``objective Bayesianism'' in his view refers mainly to priors that reflect
a lack of subjective information. These can be useful also for subjectivists,
as a ``reference''. \cite{HowUrb06} on the other hand state that the hope
for objective priors is misplaced, but that the probability axioms provide
an objective basis for logical reasoning with uncertainties. They refer to
\cite{Cox46}, who showed that probability calculus including 
Bayesian updating is implied by a certain set of axioms for such
logical reasoning, regardless of the prior. This is largely in line with
the subjectivists.

The controversy between subjectivist and objectivist proponents of epistemic
probability highlights a more general issue of statistics and even science as
a whole, namely the tension between the ideal of objectivity, and the 
impossibility to achieve it, in its idealised form, in practice. 
\cite{GelHen17} argue that instead of the ultimately futile demand of 
objectivity it is more constructive to aim at a list of virtues that constitute
aspects of objectivity, namely transparency, enabling consensus, impartiality, 
stability, and
correspondence with observable reality, acknowledging that there are also
virtues connected to subjectivity, namely context-dependence, and 
acknowledgement of multiple perspectives. Regarding epistemic probabilities,
the objectivist focus is on enabling consensus and impartiality, whereas
the subjectivist focus is on context-dependence and potentially multiple
perspectives. In order to convince others and potentially a whole scientific
community of their analysis, subjectivists
need to motivate and justify their prior choices transparently just as well
as the objectivists, despite acknowledging that others may legitimately choose
another prior.

A general issue with epistemic probabilities is that the prior probability 
assignments determine the set of potential updates. This means in particular 
that certain
prior choices that are in conflict with later observed data can never
be corrected, at least not within the Bayesian framework. A Bayesian agent
needs to explicitly assign nonzero probability to whatever seems possible to 
them, because otherwise it will remain modelled as ``impossible'' forever.
For example, if in
a Bayesian analysis the prior involves the assumption that certain future
events are \hbindex{exchangeable}, meaning that probabilities not depend on the order of outcomes, any \hbindex{posterior} 
obtained by Bayesian updating from data
will still fulfill this assumption, even if the data indicate that the 
assumption may not be appropriate. This does not seem to be a reasonable 
point of view of an epistemic agent. Consider an example regarding 
binary sequences. Even if initially it may seem to an epistemic agent 
that all possible sequences in the future 
with, say, 500 ones and 500 zeroes are equally 
likely, it seems advisable to question her own assumption and potentially
to change her mind in some cases. Imagine that 100 
ones in a row are observed, then 150 zeroes, and then a few more ones, say 15. 
This seems to show strong evidence of positive dependence between 
neighbouring outcomes, and most would now think that the outcome ``one''
should be far more likely in the next go, but having committed 
themselves to exchangeability, the epistemic agent is forced to ignore this.

It may be demanded, for this reason, that priors should not assume 
exchangeability. Furthermore, a similar case can be made against most 
regularity constraints
that may be involved in non-exchangeable models. In consequence,
the choice of priors would be far more complicated if not practically 
impossible, and much of the 
mathematical and algorithmic Bayesian machinery for computing posterior
distributions, i.e., the probabilities resulting from updating the prior
with the data, could no longer be used. Note also that a famous argument
in favour of Bayesian epistemic probability states that the prior choice  
``wears out'', i.e., the posterior distribution will ultimately be dominated 
by the data, and will in particular converge toward a ``true aleatory 
probability'' of an event in case that this exists. This is often stated 
to suggest 
that epistemic probability is just as useful as aleatory probability when it
comes to correspondence to observed reality, despite locating
probabilities in the agent rather than in reality, but the argument assumes 
exchangeability, among other things (even though assumptions can be relaxed 
a bit, see \cite[Ch. 6, 9]{DiaSky18}).

Arguably, this means that pretty much all
prior assignments of epistemic probability that are done in practice are
``wrong'' in the sense that they do not
really reflect the full belief or state of information. On top of this, 
it is well known that human agents are often biased when it comes to specifying
probabilities and assessing the implications of their own choices \citep{KST82}.

Ultimately the same that has been stated for the
aleatory interpretation already
can also be said
for the epistemic interpretation of probability. Epistemic
probabilities, be they subjective or supposedly objective, 
should not be taken as a ``really true'' reflection of the 
epistemic state of the agent. They are models, and as such, they idealise and 
simplify. The later section on ``epistemic-probability-as-model''
discusses the 
consequences of
explicitly acknowledging the model status of epistemic probabilities,

\subsubsection{Postscript on the controversy about interpretations of 
probability}\label{sps}

A lot of work \citep{deFinetti74,Jaynes03,HowUrb06,DiaSky18} suggests that
epistemic probability should be used universally, and aleatory probability 
is flawed and/or superfluous (because of the result mentioned above that under 
certain prior assignments aleatory probabilities, if they exist, can be
recovered by epistemic inference given enough data).

In particular, \cite{deFinetti74} stated that
\begin{quotation}
Probability does not exist.
\end{quotation}
This refers to the existence of objective aleatory probabilities in the world, 
and was meant as an argument that the epistemic interpretation of probability
is the only legitimate one. But I have argued above that the existence of
``true'' epistemic probabilities is just as dubious. Furthermore, the 
assumptions that allow  aleatory probabilities to be emulated in the epistemic
framework may be convenient and in widespread use, but are not necessarily 
convincing. There seems to be a tendency among authors who favour epistemic
probability to elaborate very clearly the deviation from reality of the aleatory concept taken literally, but to 
deny that the
same issue exists for epistemic probability. If the latter is acknowledged,
one may wonder whether it is worthwhile to model an epistemic agent rather
than reality itself,
in cases in which researchers want to make statements about
observer-independent reality in the first place.

\cite{Senn11} raises the issue that Bayesian analyses in practice rarely
are in line with how epistemic probability should ideally be handled, 
particularly when it comes to specifying the prior, and suggests (with
arguments similar to mine above) that this may
often be impossible. As other authors (e.g., 
\cite{Carnap50,CosToo96,Gillies00}), he acknowledges that both epistemic and
aleatory probabilities may be needed.

Defence of aleatory probability and criticism of epistemic 
probability can also be found in the literature (e.g., \cite{Mayo18}), 
but there is more general
acknowledgement that aleatory probability is not suitable for all applications
(particularly if there is nothing that can be convincingly thought of as 
repetition), and that there is a role for epistemic probability, if (as some
authors think) limited. 

I do not think
that the adoption of one interpretation of probability in one situation 
should preclude a statistician from using another interpretation in another
situation. 
A statistician can be a frequentist when analysing a randomised trial 
to compare a new drug with a placebo, and adopt an epistemic approach 
when
computing a probability that an accused is really the murderer, and in this 
sense she can be a \hbindex{pluralist}. However, if in a single analysis the implied 
interpretation of probability is not clear, results in terms of probability do
not have a clear meaning.

An interesting historical note is that a distinction between aleatory and 
epistemic probabilities was mentioned for the first time
only around 1840 \citep{Hacking75},
more than a century after the first systematic treatment of probabilities
by \cite{Bernoulli1713}. Problems associated with games of chance were among
the first sources of probability considerations, and originally fair betting,
\hbindex{classical probability} (i.e., based on counting possibilities that were 
thought of as equiprobably), and expected frequencies seem to have been thought
of as synonymous (in fact the first versions of the law of large numbers
were taken as some kind of proof of the correspondence between classical 
probability and limiting relative frequencies). There had already been 
considerations of situations that could
hardly been reduced to counting equiprobable possibilities, and 
\cite{Bayes1763} did not require equiprobability anymore, but his work seems to
allude to both epistemic and aleatory ideas. \cite{Laplace1814} defines
probabilities more explicitly as epistemic, but his treatment of measurement
error distributions looks rather aleatory, and no explicit distinction can be
found in his work.

In any case it is not an accident that apparently different concepts such as
epistemic and aleatory probabilities come with the same name and can be handled
with the same calculus, as they draw from the same sources. This may feed the 
idea that only one interpretation should be needed, which
exists particularly in the epistemic camp.

\subsection{Approaches to statistical inference}
In much literature it is suggested that adopting an either aleatory or 
epistemic interpretation of probability implies how to do 
\hbindex{statistical inference}. Indeed many (though not all) versions of the 
epistemic interpretation imply Bayesian inference, whereas classical
techniques as hypothesis tests and confidence sets have been devised
with an aleatory probability interpretation in mind. There are exceptions to 
these associations though, and I believe that it clarifies the discussion to
distinguish between interpretations of probability and approaches to
statistical inference. In particular, there is much criticism of the logic 
implied by hypothesis tests and confidence intervals that does not concern
the aleatory probability interpretation as such, and Bayesian inference is often
applied having implicitly or explicitly an aleatory interpretation in mind. 

These two approaches are the main focus here, but there are some others, see
below.

\subsubsection{Compatibility logic}
\hbindex{Hypothesis tests} and \hbindex{confidence sets} 
concern parameters of statistical 
models, or, in the \hbindex{nonparametric} 
case, certain classes of models that are
of interest in a real situation. A model parameter can for example refer to a 
physical constant that is assumed to be measured with a random error. 
Comparing the outcome of a therapy on an ordinal well-being scale to a placebo
therapy, a nonparametric set of distributions of interest may model the
distribution of results in the therapy group as the same, or alternatively as 
systematically larger than in the placebo group. In the following, \hbindex{parametric inference} will be discussed.

Data are assumed to have been generated according to such a model allowing for 
an often infinite set of parameters, and it is assumed that there is one true 
value of it.
Inference about this true value 
is made based on with which parameter values data look incompatible,
i.e., are in a suitable region with low probability. 
This region is usually
defined by a statistic that formalises a ``distance'' of the data to what is 
expected given a certain parameter value. If then the probability (``\hbindex{p-value}'') is very low that a dataset
generated from the model with the parameter value in question is as far as or 
farther away from this expectation than the observed data, the data are
deemed incompatible with the parameter value. 
In that case, a test will 
reject a prespecified
parameter value (or a set of such values) of interest. 
A confidence set will
collect all \hbindex{compatible} parameter values. Test and confidence sets have been 
criticised for simplifying binary decisions regarding compatibility, and in
fact thresholding probabilities to make such decisions can only be arbitrary.
p-values give continuous information about the degree of incompatibility
of the data with the model.

Tests and confidence intervals have been very controversially discussed recently. The centrality of the concept of compatibility can be seen in the 2016 ASA (American Statistical Association) Statement on p-values (\cite{WasLaz16}), which has a single positive
message on p-values, besides a number of negative ones: 
\begin{quotation}
p-values can indicate how incompatible the data are with a specified 
statistical model.
\end{quotation}
It is very important to understand that being ``compatible'' does not imply 
truth. Particularly, alternative models are not ruled out by any means. There
is always an infinite variety of models and usually also parameter values 
compatible with any dataset.  
Much misinterpretation of tests and confidence 
sets ignores this. Confidence sets formalise 
uncertainty in the sense that they show a range of compatible parameters, but
on top of the uncertainty within the assumed model there is model 
uncertainty that needs to be addressed in different ways.

Compatibility as defined here is not absolute, but relative to the involved
test statistic. A model may be compatible with data in some but not other 
respects. Data used for a simple linear regression for example can be compatible
with a model that has a regression slope zero, using the standard t-statistic
measuring how far away from zero the estimated slope is, but this may be 
due to nonlinearity, and a 
linearity test may reject that same model. Compatibility and incompatibility
apply to specific aspects of a model (potentially with a fixed parameter value)
rather than to the model as a whole.

It has often been stated that the rejection of a null hypothesis by a statistical test is a ``\hbindex{statistical falsification}'' in the sense of \cite{Popper59a}.
\cite{Mayo18} draws this connection explicitly. Popper's work on falsification
does not involve probabilities, and consequently it does not allow for a type I
error probability, i.e., a probability that a correct model is rejected. It
can be seen as problematic that a probability model can be deemed incompatible 
based on an event the possibility of which was essential part of the model.
This particularly leads to problems with multiple testing; a large enough number of tests will reject a true model with a large probability, even if this 
probability is small for every single test. But
in case of imprecise measurements modelled as involving random errors, 
erroneous falsification of theories due to random variation is certainly a
possibility. Troubles stemming from the possibility of ``falsifying'' a 
true model in the model world reflect a corresponding possibility in the
real world, and seem as such not inappropriate as extension of Popper's 
original ideas (in fact early work on statistical tests preceded Popper's 
work, e.g. \cite{Fisher22}). 

Furthermore, compatibility statements regarding models do not require any
``belief'' that a true aleatory probability distribution exists or even 
follows the \hbindex{model assumptions} of a test or confidence set. The model is a 
thought construct, and the thought construct can be seen as reasonable and 
helpful if the data are compatible (there may be further requirements), 
even acknowledging that the model is 
not true (the same point has been made by \cite{GreRaf19}). 
There are pitfalls though; a very large dataset can allow
so precise statements that a parameter value might be deemed incompatible 
even though it is, in terms of its real world implications, 
so close to compatible values
that it can be seen as more or less equivalent. A parameter 
might be ``rejected'' 
that implies that
a certain therapy does not improve on a placebo, even if another parameter
is still compatible with the data according to which
the improvement is
very weak and virtually useless. 

Compatibility logic is normally applied together with 
aleatory probability, but an epistemic agent may as well be interested if
her model is compatible with the observed data. This has been operationalised
by posterior predictive checks \citep{Box80,GMS96} that basically test
whether the observed data are in line with the epistemic predictive distribution, once more relative to a chosen test statistic. 

\subsubsection{Inverse probability logic}
Bayesian \hbindex{inverse probability logic} is the major alternative to compatibility
logic regarding the assessment of
probability models in the light of the data. To fix terminology, in a standard
Bayesian analysis there is usually a \hbindex{sampling model}, i.e., a distribution of the
data given a parameter, and a prior distribution of the parameter. Both of these
together are chosen prior to the analysis and can therefore be called 
``\hbindex{prior distribution}'' (as done by, e.g., \cite{deFinetti74}), even though 
in much literature this term is used for what I call 
``\hbindex{parameter prior}''.
The term ``inverse'' comes from
turning the sampling model of the data given the parameter into a 
\hbindex{posterior distribution} of the parameter given the data with help of the
parameter prior. This parameter posterior then also
allows to compute predictive distributions for future 
observations.
 
Inverse probability logic produces a probability distribution over parameters
(or models; in the following I will only mention parameters). Whereas
assessing a parameter 
as compatible with the data does not
exclude or reduce the compatibility of other parameters, 
having a probability distribution implies that parameters compete for 
probability mass. A larger probability of certain regions of the parameter space
implies a lower probability elsewhere. The probability axioms, whether 
interpreted in an epistemic or in an aleatory manner, imply that there is only 
one truth (only one of two disjunct events can obtain), even though not 
known with certainty.

Often posterior distributions are interpreted as distributing probabilities
about where to find the true parameter value. The idea of a true
parameter value is a traditional frequentist one.
\cite{deFinetti74} 
argued that posteriors should be interpreted regarding observable quantities
such as future observations and not regarding unobservables such as true 
parameter values that may well not exist. The parameters for him are 
helpful devices to summarise prior and posterior assessments, but 
\hbindex{betting} (as crucial for the traditional subjectivist Bayesian approach)
should concern observables, so that bets can ultimately be decided.
But according to 
\cite{DiaSky18}, de Finetti's Theorem means that for a subjectivist, belief
in exchangeability or a suitable generalisation of it implies belief in
the existence of limiting relative frequencies, and therefore a limiting
probability distribution that can be parameterised. Applied Bayesians
sometimes argue that even though true aleatory distributions might not exist,
in many cases the parameter of interest
can be interpreted as a meaningful and potentially
objectively existing quantity such as a physical constant that can only be
measured with an error treated as random, giving rise to epistemic
uncertainty.

\subsubsection{Compatibility logic vs. inverse probability logic}
There are advantages and disadvantages of both approaches. 
Many Bayesians such as 
\cite{DiaSky18} have pointed out that p-values and confidence
levels are regularly misinterpreted as probabilities regarding the true 
parameter, because these should be the ultimate quantities of interest in
statistical inference, or so it is claimed. 
Inverse probability logic deals with combining different inferences such as
multiple testing, which creates trouble for standard compatibility logic 
approaches, in a 
unified and coherent way.
Frequentists are not only interested in compatibility, but also in estimation;
finding a  best model amounts to a competition between models. A 
Bayesian can argue that in this case a probability 
distribution over parameters provides more differentiated and appropriate 
information about how the
parameters compare and to what extent one parameter value 
is more likely than others.
Using \hbindex{confidence distributions} (\cite{XieSin13}), which define a probability
distribution over the parameters based on the confidence levels at which 
parameters are still in an appropriately defined confidence set,   
frequentists involving compatibility logic 
can give more detailed information about how parameters compare 
as well, but Bayesians hold that a proper probability distribution is more
intuitive and less prone to misinterpretation. Confidence distributions
are strongly connected to \hbindex{fiducial probabilities}, 
which  \cite{Fisher36}
defined in order to make probability statements about parameters without
requiring a prior. This implies some kind of inverse probability logic, even
though Fisher distanced himself from the Bayesians. Modern work on fiducial
inference \citep{HILL16} uses fiducial distributions of parameters as 
inferential tool allowing for compatibility statements rather than as actual
epistemic distribution alternative to the Bayesian approach.

By allowing many models to be compatible with the data
at the same time, 
compatibility logic is more obviously in line with the
attitude that models are idealisations and not really true.
\cite{Davies95} argued that if many models are valid approximations for the
same real situation, inverse probability logic is 
inappropriate, because if for example ${\cal N}(0,1)$ is a reasonable 
approximation of the truth (${\cal N}$ denoting the Gaussian or normal 
distribution), ${\cal N}(10^{-10},1)$ is a reasonable approximation
as well, whereas according to inverse probability logic any two models compete
for a part of the unit probability mass.

Furthermore, it may seem unfair to criticise tests and confidence intervals 
based on misinterpretations. Arguably many users do not only want to know the
probability for certain parameter values to be true, which indeed tempts 
them to 
misinterpret confidence levels and p-values. But arguably they also want this
probability to be objective and independent of prior probability
assessments, which to make
up they have a hard time. This combination
is not licensed by any properly understood philosophy of
statistics, and ultimately statisticians need to accept that their job is
often not to give the users what they want, but rather to defy wrong 
expectations.  

The role of the prior distribution in inverse probability logic 
is a major distinction between the two approaches. Bayesians argue that 
the prior 
is a good and very useful vehicle to incorporate prior information.
Actually prior information enters frequentist modelling as well (see the later
discussion of frequentism-as-model), 
but the Bayesian prior is still an additional tool on top of
the options that frequentists have to involve information.
But the requirement to set up a prior can also be seen as a major problem with
inverse probability logic, given that prior information does not normally come
in the form of prior probabilities, and that it is actually in most cases very
difficult to translate existing information 
into the required form. It is not an accident that 
a very large number of applied Bayesian publications come with no or very 
scarce subject matter justification of the prior, and in most cases the prior
information is very clearly compatible with many different potential 
priors, with comprehensive
\hbindex{sensitivity analysis} rarely done.
If the sample size is large enough for the prior to lose 
most of 
its influence, one may wonder why to bother having one. 
The question whether there is prior information that is meant to
have an impact on the analysis and can be
encoded convincingly in the form of a prior distribution is a key issue for
deciding whether inverse probability or compatibility logic will be more 
promising in a given application of statistics. 

\subsubsection{Prediction quality}\label{spredict}
Given the problems connecting aleatory and epistemic probabilities to any 
objectively existing truth, a pragmatic stance may hold that inference about 
``true parameters'' is not of interest in its own right. Rather, probability
modelling is worthwhile to the extent that it enables the user to deal with
uncertainty about the future, and more precisely, \hbindex{prediction} of future 
observations. This point of view is dominant in the machine learning community,
and becomes more popular also in the statistical community, following 
\cite{Breiman01}, who pointed out that traditional statistical modelling (be it
aleatory or epistemic) often produces an inferior prediction performance 
compared to approaches that aim more directly at optimising \hbindex{prediction
quality}. Such approaches may not even involve any probability modelling, or
only in such a convoluted way that clear interpretations regarding the 
underlying workings of reality leading to the predictions are hardly possible. 
Prediction quality can be assessed using data driven techniques such as 
\hbindex{cross-validation} and \hbindex{bootstrap} 
that apparently do not refer to specific 
probability models. A worry is that performance achieved on specific training, 
test, and validation data may not necessarily generalise to future data.
Probability models, usually nonparametric 
and interpreted in an aleatory 
manner, are used to theoretically underwrite and investigate 
prediction algorithms as well as validation techniques (e.g., 
\cite{SchSmo02,BaHaTi21}).

I agree with Breiman that 
probability modelling is certainly not mandatory, which 
runs counter to the naive idea that finding the true model is the 
ultimate aim of statistics. On the other hand, \cite{Breiman01} and much work 
from the machine learning community seem to reduce aims of modelling to 
prediction quality as all-dominating issue, which neglects the role of models
for communication and building understanding (see the section on 
mathematical models and reality),
even though it is probably valid 
to state that prediction quality is very often at
least implied by the aim of analysis. Recently this issue has been
acknowledged also in the machine learning community, and there is a raising 
number of publications on the \hbindex{interpretability} of the outcomes of
machine learning algorithms 
\citep{RCCHSZ22}.

Probability models allow for a quantification of uncertainty 
that is often
desired in prediction problems. Bayesian approaches deliver such \hbindex{uncertainty quantification} automatically, of course relying on the appropriateness of
the models. This is harder for
machine learning techniques that are not based on probability models, 
but work exists, e.g., \cite{LaPrBl17}.

Also there are doubts to what extent prediction quality for future events
can be reliably assessed.
For example, there is the possibility
that the process of interest in the future may change, and a transparent model
with solid subject matter justification may be more easily adapted, as 
\cite{Hand06} argues. Cross-validation and other methods for 
assessing prediction quality imply independently and identically distributed 
(\hbindex{i.i.d.}) repetition, which may be problematic. 

\subsubsection{Other approaches}
Many further approaches to statistical inference exist. Technically, 
parameter \hbindex{estimation}, i.e., finding a best guess, 
does not fall into any of the earlier listed categories. Estimation can be of 
interest regarding both aleatory and epistemic probabilities; in the latter case
particularly where the parameter is identified with a not directly observable
quantity that is assumed as existing and meaningful, e.g., in measurement 
problems with measurement error. Estimators are usually solutions of optimality
problems, although there are also qualitative criteria for estimators such
as consistency, or robustness against model misspecification. Statisticians 
often insist that estimators should come with an indication of their 
uncertainty, which usually then requires either inverse probability or 
confidence sets (which may involve nonparametric bootstrap techniques).

Some authors \citep{Hacking65,Edwards72} advocated the use of the 
\hbindex{likelihood}
for assessing the support that data lend to certain models, without
involving prior distributions. Likelihoods are hard to compare particularly 
over models with different complexities, but information theoretic approaches
such as \hbindex{Akaike's Information Criterion} 
can be used for such comparisons 
\citep{BurAnd02}. 

A rich literature exists regarding \hbindex{imprecise probabilities}, 
which combine 
modelling uncertainty about the occurrence of an event $A$ with the strength
of knowledge regarding $A$. Lower and upper probabilities are defined, with the
interpretation that generally weak knowledge is expressed by a big difference
between the two. This can be interpreted in an epistemic or aleatory manner,
referring either to uncertainty of an agent, or to upper and lower bounds 
of probabilities in a set of aleatory probability models \citep{Walley91}.
The latter is related to \hbindex{robust statistics}, 
which aims at good frequentist 
properties of estimators, tests, and confidence sets given a nominal model,
and stability in a suitable neighbouring class of models.
\hbindex{Dempster-Shafer belief functions} \citep{Shafer76} 
are another variant of
considering upper and lower plausibility indications. \cite{ShaVov19} propose
new foundations of probability and inference based on specific \hbindex{betting} games.

Less formally, visual and \hbindex{exploratory data analysis} 
also deserve a mention,
providing sometimes valuable alternatives to model-based inference. This may
be better to appreciate from a philosophical position that does not grant
mathematical formalism a generally superior access to reality. 

\subsection{Mathematical models and reality}
\label{smodel}
I wrote \cite{Hennig10} mainly out of the conviction that the debate about the
foundations of statistics often suffers from the lack of a general account of 
\hbindex{mathematical models} and their relation to reality.  

On the frequentist side often a rather naive
connection between the models and \hbindex{reality} is postulated, which is then easily 
criticised by advocates of epistemic probability, implying that
because a naive realist idea of frequentist probabilities is not convincing,
probability should better be about human uncertainty
rather than directly about the reality of interest. 
But modelling human 
uncertainty is still modelling, and similarly naive ideas of how such models
relate to ``real'' human uncertainty are problematic in similar ways.

Inspired by radical and social \hbindex{constructivist philosophy} 
\citep{vonGlasersfeld95,Gergen99}, 
I have argued that mathematical models are
thought constructs and means of communication about reality. I have 
distinguished ``observer-independent (or objective) reality'', 
``\hbindex{personal reality}'' (the view 
of reality constructed by an individual observer and basis for her actions),
and ``\hbindex{social reality}'' (the view of reality constructed by communication between
observers). Taking a constructivist perspective, I have argued that 
\hbindex{observer-independent reality} 
is not accessible directly to human observers, 
and that, if we want to talk about reality in a meaningful way, it makes sense 
to refer to personal and social reality, which are accessible by the 
individual, a social group, respectively, rather than only to 
observer-independent reality. Observer-independent reality manifests itself 
primarily in the generally shared experience that neither an individual observer
nor a social system can construct their reality however they want; we all
face resistance from reality. This 
resonates with \cite{Chang12}'s ``\hbindex{Active Scientific Realism}'' 
(see also 
\cite{Chang22}): 
\begin{quotation}
I take reality as
whatever is not subject to one's will, and knowledge as an ability
to act without being frustrated by resistance from reality. This
perspective allows an optimistic rendition of the pessimistic induction,
which celebrates the fact that we can be successful in
science without even knowing the truth. The standard realist
argument from success to truth is shown to be ill-defined and
flawed.
\end{quotation}

In \cite{Hennig10} I interpret 
\hbindex{mathematical models} as a particular form of social reality, based on
the idea that mathematics is a particular form of communication that aims at 
being free of ambiguity, and that enables absolute agreement by means of 
proofs. 
Science is seen as a social endeavour aiming at general agreement about 
aspects of the world in free exchange,
for which unambiguous mathematical communication should
be obviously useful.
This comes to the price that mathematics needs to unify
potentially diverging personal and social observations and constructions.
There is no guarantee that by 
doing so mathematical models come closer to observer-independent reality; in 
fact, this view sees the domain of mathematics as distinct from the domain of 
observer-independent reality. The connection is that the constructs of personal
and social reality are affected by ``perturbation'' (or, as Chang puts it, 
``resistance'') from observer-independent 
reality, and it can be hoped that such perturbation, if observed in 
sufficiently related ways by different individuals and social systems, can be
represented in mathematical modelling, which then allows to analyse the 
represented reality using the machinery of mathematics and logic. 
There is also 
repercussion of communication, including mathematics, on social and personal
constructions of reality. Mathematical modelling has an effect on 
the world view and dealings with the world
of social systems and individuals. We can hope for this effect to be 
beneficial, but this may not always be the case, because important 
subtleties and differences between individual and social 
realities may be lost in mathematics. See, e.g., the criticism of the
unreflected use of quantification in psychology \citep{TaSlNe16} and a more 
general exposition of the role and history of quantification \citep{Porter96}. 

According to constructivism, it should not be hoped that mathematical 
models represent 
reality ``correctly'', but they can have many uses anyway. They can
represent important aspects of 
the modeller's perception of reality. 
Accordingly, a major benefit that they
can have is to communicate the modeller's view unambiguously, and in this
way to support agreement (or potentially ``agreement to disagree''). 
They can
explore different scenarios, reduce complexity, help decision making, and
enable predictions (that can be 
checked against observed reality). Sometimes the best way to learn from a
model is to understand in which way it fails.

Another important benefit is the stimulation
of creativity and method development. 
A standard way to solve a data analytic problem is to set up
a probability model, and to derive inference methods that are in some formal 
sense justified or even optimal given that model. This does not rely on the 
truth of the model, but of course if the model is inappropriate, results
can be misleading. Constructivism does not advocate that ``anything goes'';  
the modeller needs to convince themselves and her
community that the model is a good basis for inference, which can involve
both checking against observations, and explaining how existing knowledge and 
thought is related to the model. Rather than licensing an arbitrary use
of models, the focus is on the responsibility of the modeller to
explain and defend her modelling choices.
Widely accepted mathematical models in
science have been discussed, questioned, defended, and possibly updated, 
and have been found to be in agreement with observations in lots of situations.

The constructivism I adhere to may be these days
rather unpopular in philosophy, and what Chang refers to as ``standard realism''
in his quote above has won some ground. However, given the rather tedious 
controversy that philosophers have about the objective 
``existence'' of long run frequentist or single-event chances and propensities
(see \cite{Eagle19}), it seems as attractive as ever to treat them 
as a thought construct and to look at how they are used and how they relate 
to observations in order to circumvent
the enormous amount of problems that come 
with postulating the objective ``existence'' of such a thing.

I acknowledge that there is a lot of philosophical work on models in science
that is not naive realistic, and concerned in more detail with how modelling
actually works, see \cite{FriNgu20}. Much of this is compatible with the view
taken here, as also a constructivist view implies that models represent reality,
reality referring here to the perceived reality of the modeller and possibly 
her social system.

\subsubsection{Reality, observations, operations}
In order to make a convincing case for a proposed model, it needs to be
connected with reality, or more precisely, with the reality perceptions of
the people in the community at which the model is addressed, at an individual
and social level. Von Mises, Reichenbach (another contributor to the
foundations of frequentism),
de Finetti, and Ramsey, were all influenced by \hbindex{logical empiricist} 
ideas, 
and by the \hbindex{operationalist} view of \hbindex{measurement} 
\citep{Bridgman27,Creath21,Gillies00,Galavotti05} when proposing their
foundations of aleatory and subjectivist epistemic probability, respectively.
They
built their concepts on well defined evaluation operations referring to
observations, and tried to avoid \hbindex{metaphysical} 
concepts (even though von
Mises' limiting relative frequency has been charged as metaphysical).
Von Mises
defined probabilities as what is measured, with increasing precision, by
observed relative frequencies; de Finetti detailed the betting operations
necessary for eliciting prior belief, and declared that the resulting bets
should concern future observations rather than unobservable parameters.

Measurement is central to the philosophy of mathematical models outlined above,
as measurement is the transformation of informal perceptions to mathematical
objects that can then be handled by mathematical models and methodology. 

Logical empiricism and operationalism have been criticised in various ways
and are no longer very popular \citep{Creath21,Chang21}, but
\cite{Chang21} argues that operationalism is still relevant regarding central
issues in the philosophy of science, 
and \cite{VanFraassen80} proposed a more modern
version of empiricism, sharing its anti-metaphysical 
(sometimes branded ``anti-realist'') commitment. From my constructivist 
perspective, early empiricists and operationalists took a too extreme and 
reductionist point of view by denying the meaning of thought constructs that
could not directly be tied to operations and observations. Operations and 
observations rely themselves on theories. It is important to acknowledge that
measurement operations and outcomes are not an objective ``given'', and that
the distinction between a measurement and what is measured is an elementary part
of scientific reasoning. Scientists need to argue the validity of their 
measurement operations regarding what they want to achieve, and observations
can be criticised meaningfully as potentially erroneous. This does in my view
not refer to the inaccessible observer-independent reality, but rather to 
the thought constructs of the scientists and their community.

But I agree with Chang and Van Fraassen that
the focus on observations and operations is generally helpful for clarifying
meaning, validity, and implications of scientific concepts. Observations and 
operations do not have absolute authority and need to be negotiated and agreed,
but they seem much better suited to enable clear negotiations and agreement 
(or clear disagreement) than reference to an unobservable ``truth''. 
Particularly regarding epistemic probability, \hbindex{betting schemes} give a much 
clearer idea what it means to state that ``the probability of an event is 0.4''
than referring to the rather abstract setup of \hbindex{Cox's Theorem} 
as motivation why reasoning about plausibilities
can be formalised by probabilities. For frequentism, von Mises's concept of 
the ``\hbindex{Kollektiv}'' clarifies that aleatory probabilities need to be 
defined relative to a ``\hbindex{reference class}'' of experiments that qualify
to be counted, which is often ignored in practical data analysis but can be
a source of trouble \citep{Hajek09}.

In this vein the contributions of von Mises, de Finetti, and others, can be 
re-appreciated despite the criticism that they attracted when interpreting their
work as attempt to convince the community of the ``truth'' or 
``objectivity'' of these concepts. Without this ambition, what they 
achieved is a thorough elaboration of the (partly problematic) 
implications and the operational and 
empirical meaning of the corresponding probability interpretations. My position
is not that models can be arbitrarily chosen, but rather that the modeller 
needs to be clear about how the involved probabilities are to be understood, and
to what extent they can be empirically tested, or used in empirical work. This 
will ultimately contribute to transparent communication, and to
stability and reliability of the outcomes.

\section{Frequentism-as-model} \label{sfam}
The frequentist (aleatory) interpretation of probability and 
frequentist inference such
as hypothesis tests and confidence intervals have been strongly criticised
recently (e.g., \cite{Hajek09,DiaSky18,WSL19}). In applied statistics they are
still in very widespread use, and in theoretical statistics the number of 
still appearing new works based on frequentist principles is probably not 
lower than that of any competing approach to statistics.

Even defenders of frequentist statistics such as \cite{Mayo18} admit that
frequentist inference is often misinterpreted, while arguing that such
misinterpretation is a major source of unfair criticism of frequentism. By and
large I agree with this. I will argue that what I call ``\hbindex{frequentism-as-model}''
avoids many issues that 
critics have raised, and allows for a better understanding of how results of 
frequentist inference can be interpreted. 
    
Here are some influences and important ideas from the literature. I already 
mentioned
the ``long run \hbindex{propensity}'' interpretation of probability by 
\cite[Ch. 7]{Gillies00}. 
Although his outlook is more traditionally realist than mine, 
I see this as an insightful interpretation of  
how frequentist statisticians actually use their models. 

\cite{Tukey97} re-interpreted the role of frequentist models in statistical
inference as ``challenges'' that procedures have to face in order to show
that they are worthwhile in a given situation. He recommended using 
\begin{quotation}
bouquets
of alternative challenges
\end{quotation}
to investigate how well a procedure may work
in a real situation, including worst cases, rather than finding an optimal 
procedure for a single model.
He wrote 
\begin{quotation}
(more honest foundations of data analysis) have to include not assuming that we always know what in fact we never know -- the exact probability structure involved (\ldots)
almost nothing is harder than verifying a stochastic assumption, something I find hard to believe has ever been done in even a single instance.
\end{quotation}
Many different models can be compatible with the same situation, and it is often misleading to base an analysis on just one model.

\cite{Davies14}'s view of statistical models is similar: 
\begin{quotation}
The problem is how to relate a given data set and 
a given model. Models in statistics are, with some exceptions, ad hoc, 
although not arbitrary, and so should be treated consistently as 
approximations and not as truth. I believe there is a true but unknown 
amount of copper in a sample of water, but I do not believe that there is
a true but unknown parameter value. Nor am I prepared to behave as if 
there were a ‘true but unknown parameter value’, here put in inverted 
commas as statisticians often do to demonstrate that they do not actually 
believe that there is such a thing. Moreover the model has to be an 
approximation to the data and not to some ‘true but
unknown generating mechanism’. There may well be some physical truth behind 
the data but this will be many orders of magnitude more complex than anything 
that can be
reasonably modelled by a probability measure. Furthermore there will be a 
different truth behind every single observation. (\ldots) The definition of 
approximation used in this book is that a model is an \hbindex{adequate
approximation} to a data set if ‘typical’ data generated under the model 
‘look like’ the real data.
\end{quotation}
Again this implies that many different models
can be adequate approximations of the same data, and that the job of a model
is not to be true. 



A difference between Davies's and my own view is that I think that ``treating
reality {\it as if} it behaves as the model says'' (temporarily, and with
skepticism) is a valid and helpful description of the handling of
models, whereas Davies thinks it is enough to define the 
probability models' ability to 
approximate data mathematically, and no further interpretation is required.
\cite{DavDum22} give an example for a nonstandard use of probability
models that uses them as calibration device rather than ``assuming'' them to
be real in any way.

An aspect of the work of both Tukey and Davies is an emphasis on 
\hbindex{procedures} for data analysis rather than models. The latter may 
be used to
assess and potentially define or inspire procedures, but play just an 
auxiliary role, and are often not required. \cite{Tukey93} wrote
\begin{quotation}
(\ldots) once we have ceased to give a
model’s truth a special role, we cannot allow it to ``prescribe'' a procedure. 
What we really need to do is to choose a procedure, something we can be 
helped in by a knowledge of
the behavior of alternative procedures when various models (i.e., various challenges) apply (\ldots)
\end{quotation} 
I sympathise with this attitude, even though the present
chapter on models is not the right place to discuss it in detail. In my 
impression, Tukey and Davies treat the models as \hbindex{data generating processes}, and
as such they are interpreted as aleatory.
 

I will first define what I mean by frequentism-as-model. Then, in order 
to
understand its implications, the role of i.i.d. models and subject matter
information is discussed.
Reasons why frequentism-as-model can be useful despite
the separation between reality and model are given.
I will then treat statistical inference based on 
frequentism-as-model, including the role of model assumptions, robustness and
stability, and the key concept of ``interpretative equivalence''.
The last subsection concerns the use of inverse probability logic with
aleatory probabilities in the spirit of
frequentism-as-model.

\subsection{Frequentism-as-model as interpretation of probability}\label{sfamip}

\hbindex{Frequentism-as-model} 
is a version of the aleatory interpretation of probability
and can be seen as a flavour of frequentism in the sense that a
probability of an event $A$ refers to the limiting relative frequency for 
observing $A$ imagining an
idealised infinite \hbindex{repetition} of a random experiment with
possible outcome $A$. This comes with problems, as discussed earlier.

Von Mises and most other frequentists claimed that their 
interpretation is applicable 
to experiments allowing for lots of (idealised infinitely many, idealised 
identical) repetitions, 
but in fact in standard frequentist probability modelling such repetitions are
modelled as \hbindex{i.i.d.}, meaning that 
a whole sequence of repetitions is modelled by a single probability model, 
modelling probabilities not only for the outcomes of a single replicate, but 
for combinations of outcomes of some or all replicates. Central results of
probability theory such as the laws of large numbers and the central limit 
theorem apply to such models. Applying the traditional
frequentist interpretation of probability to them would require whole 
independent and potentially infinite sequences to be repeated, which of course
is impossible in reality. 

Frequentism-as-model deals with the issues in two ways. Firstly, it emphasises
that probability is a model and as such fundamentally different from observed 
reality. Adopting frequentism-as-model means to think about an experiment or a 
situation {\it as if} it were generated by a frequentist process, temporarily,
without implying that
this really is the case. This means that it is not required to 
make a case that the experiment is really infinitely or even very often 
repeatable and will really lead to the implied relative frequencies. 
Particularly this means that all available observations can be modelled as
i.i.d. without having to postulate that the whole sequence of observations can
be infinitely or even just once repeated. Frequentism-as-model does not imply
a belief in objective randomness and is also compatible with determinism.
Obviously, 
insisting on the fundamental difference 
between model and reality in this way
raises the question how we can use such a model to learn
something useful about reality, which I will address below. 
Von Mises and other frequentists already acknowledged that frequentist
probabilities are idealisations. To me it seems, however, that in traditional
frequentism this is only
a response to critics, and that otherwise it does not have consequences 
regarding data analyses and their interpretations. 

Secondly, whereas it is not clear how to connect the limits of traditional 
frequentist infinite sequences to an observed finite amount of real data as
any limit is invariant against arbitrary changes of any finite beginning of 
a sequence, i.i.d. models, as well as models with other fully 
specified dependence
and non-identity structures, make probability statements about finite sequences
of observations, and these can be checked against the actual observations. It 
can be found out if and in which sense the observed sequence would have been
``typical'' according to the specified model. In fact, whereas we may not 
be able to observe 
another real data sequence of the same length, and surely not arbitrarily many,
we are able to generate an almost arbitrary number of repetitions of sequences
from the model involving just the light idealisation that we would need a 
perfect random number generator. Even without a random number generator, 
in many situations probability
theory can tell us what to expect if the model holds. This resonates with 
Gillies's
long run propensity interpretation of probability, where 
probabilities are not 
defined referring to infinite sequences, but rather as tendencies to observe in
finite experiments a ``typical'' (i.e., large probability) outcome as specified
by the model. 

An issue with this is the definition of ``typical''. For example, assuming an
i.i.d Bernoulli model, a sequence that has 50 zeroes, then 50 ones, then 50
zeroes, than 50 ones again, and so on (called 50-50-sequence in the following),
has the same probability as a randomly 
looking sequence of zeroes and ones that gives us no intuitive reason to 
doubt the i.i.d. assumption. On what basis can we claim that the i.i.d. 
model is \hbindex{adequate} (to use Davies's terminology) for the latter sequence but not
for the 50-50-sequence? This requires subjective researcher input. 
The researcher
needs to decide about the specific way, or ways, in which a sequence has to 
be typical according to the model in order for the model to be ``adequate''. 
For example, the researcher may decide that the model is only useful for data
in which the lengths of ``runs'' of zeroes or ones in a row are not longer 
to what
is expected under the model, in which case the model will be ruled out for the
50-50-sequence (which can be formally done using the \hbindex{runs test},
\cite{Lehmann86}, p.176),
whereas the irregular sequence will count as typical.
A reason for such a decision 
may be that the researcher thinks that sequences with long runs of
the same outcome are better modelled by models involving positive dependence
between the individual binary experiments, and that under such models it can 
be shown that the method of analysis the researcher would want to apply to an 
i.i.d. Bernoulli sequence will be misleading. This could for example
be a confidence interval for the
probability of observing a one. Keeping in mind the
fundamental difference between model and reality, the researcher in this 
situation does not make a decision about which model she believes is in fact 
true. For such a belief, the subjective decision to distinguish between 
two sequences
of same individual probability may look questionable. The actual
decision is rather about 
case-dependent criteria regarding in terms of what models they want to think 
when 
analysing the data, informed by the aim of data analysis as well as the 
characteristics of potential alternative models, and knowledge about the nature
of the data and potential reasons for dependence. Note though that if the 
researcher uses the runs test as decision rule, she will never be able to 
observe a 50-50-sequence whenever she uses the i.i.d. Bernoulli model, under
which such sequences should actually be possible. This is an instance of
the ``misspecification paradox'', see the subsection on model assumptions.

Probability theory offers ways
to help the researcher making such decisions, by for example addressing 
questions of the form ``If indeed the model is as assumed, how likely is it to
decide against it; or if a different model is true, how likely is it to not
detect it and wrongly stick to the original one; and how bad would the 
consequences of such an erroneous decision be in terms of the probability of 
getting a misleading result of the final analysis?'' The researcher may also
decide to adopt a more robust final method of analysis in order to deal
appropriately with a larger number of models between which the data cannot
distinguish. 

By always keeping in mind that reality is only treated temporarily {\it as if} 
generated by a certain model, not ruling out the possibility that it could be
different (for which we may consider different models), and even taking 
this possibility into account when making further
data analytic decisions, frequentism-as-model takes the fundamental difference
between model and reality much more seriously into account than a standard
frequentist analysis, in which a model is used and normally no longer questioned
after some potential initial checks. This leads to interpretations of the final
results that often seem to naively imply that the model is true, even if the 
researcher using such an 
approach may well admit, when asked explicitly, that reality actually 
differs from the model. 

\subsection{The role of i.i.d. models and subject-matter knowledge} \label{siid}

Thinking about reality {\it as if} it were generated by a certain 
probability model has implications. A critical discussion of these 
implications is a way
to decide about whether or not to adopt a certain
model, temporarily, and on top of this it bears the potential to learn about
the real situation. If a sequence is thought of as generated by an 
\hbindex{i.i.d.}
process, it means that the individual experiments are treated as identical
and independent. De Finetti made the point that whatever can be distinguished 
is not identical. Treating experiments
as identical in frequentism-as-model does not mean that the researcher
believes that they are really identical, and neither do they have to be
really identical in order to license the application of such a model.
Rather it implies that the differences
between the individual experiments
are temporarily ignored or, equivalently, treated as irrelevant.
Analogously, using an
independence model does not mean that the researcher believes that
experiments are 
really independent in reality, but rather that she assesses potential sources 
of dependence as irrelevant to what she wants to do and how she wants to 
interpret the result. In this way, i.i.d. \hbindex{repetitions} are 
constructed by the researcher.
Of course, depending on what kind and strength
of dependence can be found in reality, this may be inappropriate. On such
grounds, the model can be criticised based on subject-matter knowledge,
and may be revised. This allows for a 
discussion about whether the model is appropriate, on top of possible checks 
against the data, which may have a low power detecting certain deviations
from an i.i.d. model.

I.i.d. models, and their Bayesian counterpart, \hbindex{exchangeability} 
models,
play an important role in frequentism-as-model as well as in
applied statistics in general. Of course many non-i.i.d. models are used such as
\hbindex{regression} 
where the response distribution is non-identical depending on the
explanatory variables, and models
for \hbindex{time series} or \hbindex{spatial dependence}. Even those models 
will normally have an i.i.d. element, such as i.i.d. residuals or innovations,
be it conditional in a Bayesian setup. This is not an accident. Treating
experiments as identical and independent (or having regular and fully
specified dependence or non-identity structures that will usually
involve an i.i.d. element) specifies how an observation from
one experiment can be used for learning about another, which is the core
aim of statistics. Statistics relies on models that allow us to use different
observations to accumulate information. This particularly means that the
statistician will need to make decisions to ignore certain potential
irregular differences or dependencies in order to be able to do her job
to aggregate information,
regardless of whether her position is frequentism-as-model, traditional
frequentism, or an epistemic Bayesian approach. Obviously i.i.d. models
do not necessarily lead to convincing results, and ultimately it is a matter
of experience and practical success in what situations the use of these models
is appropriate. Note in particular that it is possible with a
sufficiently large amount of data to distinguish, with large probability,
i.i.d. data from data with a regular \hbindex{dependence} 
structure such as ARIMA. It is however
not possible to distinguish i.i.d. data from arbitrarily irregular structures
of dependence or non-identity. A model that states that the first observation
is random but conditionally on it all further observations are fixed
with probability one can
never be ruled out based on the data alone, and subject-matter arguments are
required to make a reasonably regular model plausible.

At the stage of analysis, both checking of adequacy based on the data
and understanding and discussion of the subject matter background
are of crucial importance. The charge of an ignorance of subject-matter
information that Bayesians often raise against
frequentists does not affect frequentism-as-model. On the flipside,
frequentism-as-model obviously relies on subjective decisions of the
researchers, so that a statistician holding it cannot proudly claim that
analyses are fully objective, as some frequentists and 
\hbindex{objective Bayesians}
like to do, which is questionable anyway (\cite{GelHen17}).

As an example for subject-matter arguments,
in a much-cited and influential study that prompted a Lancet 
editorial (\cite{Lancet05}) with the title ``The end of homeopathy'',
\cite{Shangetal05} carried out a meta-analysis of eight studies comparing 
\hbindex{homeopathy} with a \hbindex{placebo} using a standard 
\hbindex{meta-analysis} model with a \hbindex{random
study effect} assumed i.i.d. (note that this implies \hbindex{dependence} between the
observations within the same study). They found that there is no evidence that 
the odds ratio between ``homeopathy works'' and ``placebo works'' is different 
from one and concluded (not inappropriately) that 
\begin{quotation}
this finding is compatible with the
notion that the clinical effects of homeopathy are placebo effects.
\end{quotation}

Here are two implications of the i.i.d. model for the random study effect.
Firstly, modelling it
as identically distributed implies that the differences
between studies are attributed to random variation, which basically means that
knowledge of known differences between the studies is ignored for modelling.
This particularly includes differences between different modes of applying
homeopathy. For example, there were only two studies used in the meta-analysis
in which homeopathy was applied in the classical way with individual 
repertorisation, which is seen as the only proper way of applying homeopathy 
by many homeopaths. Modelling the  study effect of these studies as coming from
the same distribution as all the other study effects may be justifiable if
there is an a priori belief that homeopathy basically is a placebo and the
differences between application modes are irrelevant, but it is not suitable
for convincing skeptical supporters of classical homeopathy who would be 
willing to accept a fair study.

Secondly, an expected value of 0 of the random effect
is not assumed for every single
study, but over all studies, with a potentially arbitrary large between-studies
variance. This implies that the effective sample size for 
estimating the odds ratio is not the number of individual patients 
treated in all 
studies (several thousands), but the number of studies, $n=8$, and $n=2$ for
classical homeopathy only, which the authors may use to justify their decision
to not make a modelling difference between different modes of applying 
homeopathy, selected by \cite{Shangetal05} by a study quality criterion. 
The power of the resulting test
is therefore very low,
and consequently 
the confidence interval that the authors give for the odds ratio 
is very wide. This 
argument applies to meta-analyses with a random effect generally; in 
the literature one can sometimes find the informal
remark that such analyses require a large number of studies, see
\cite{KMS08}. On top of that it could be discussed whether there may be
reasons for assuming 
systematic bias over all studies, which in the model would amount to
a nonzero expectation of the random effect.

Obviously $n=8$ is not sufficient to check the 
i.i.d.-assumption for the random study effect at any reasonable power. 
Rather than having any connection to an observable ``truth'', the 
random effect is 
a convenient modelling device allowing for potential
individual study effects to be captured by a single parameter. A low
number of parameters is, according to statistical theory, 
useful with a small $n$. This comes at the price of precluding the possibility 
to involve specific information about differences between the individual trials.
This implication is usually ignored.

Given this, it looks very dubious that the Lancet (2005)'s 
issue editorial concluded: 
\begin{quotation}
Surely the time has passed for selective analyses, 
biased reports, or further investment in research to perpetuate the 
homeopathy versus allopathy debate.
\end{quotation}
Regarding the purely
statistical evidence, the meta-analysis 
could as well be used to argue that there
are not yet enough high quality studies.


In principle such discussions can also take place regarding the traditional
\hbindex{frequentist} use of models. but frequentism-as-model re-frames these 
discussions
in a helpful way. It makes us aware of the implicit meaning of the model
assumptions, which can be used to discuss them. And it 
emphasises that what has to be decided is not
the truth of the model, but rather a balance between the capacity of the
model to enable learning from the existing observations about future
observations or underlying mechanisms on one hand, and on the other hand
taking into account
all kinds of peculiarities that may be present in the real situation of
interest, but that may be hard to incorporate in such a model despite
potentially having a more or less strong impact on the results.

A key aspect is that rather than thinking in binary terms about whether a 
model is true or not, frequentism-as-model implies that the connection 
between observed reality and models can be assessed
gradually as more or less close. It is neither impossible nor inadmissible to 
conceive a situation as a random experiment that cannot or only a very 
limited number of times be repeated, and to use a model in the sense of 
frequentism-as-model for it. This would imply to {\it think of} the situation 
as hypothetically repeatable, and of certain outcomes having a certain
tendency, or, in standard philosophical terms, propensity to happen, which 
would materialise in case of repetition, but not in reality.

Therefore frequentism-as-model does not bar \hbindex{single-case probabilities} 
as
von Mises's flavour of frequentism does. But it is more difficult for the 
person who puts up such a model to convincingly justify it compared to
a situation in which there are what is interpreted as ``replicates''.
Convincing justification is central
due to the central role of communication and agreement for science 
in the philosophy outlined above. 
Whatever the model is used for, it must be taken into account that hypothesised
values of the single-case probability cannot be checked against data. 
This 
changes if the experiment is embedded in a set of experiments that are not 
directly seen as repetitions of each other, but for which an assumption is 
made that their results are put together from systematic 
components and an error term, and the latter is interpreted as \hbindex{i.i.d.} 
repetition. This is actually done in standard frequentist \hbindex{regression} 
analyses,
in which the \hbindex{explanatory variable} or vector {\bf x} is often assumed fixed.
For a given value or vector of values of the explanatory variable(s) {\bf x},
a distribution for the response $y$ is implied, despite the fact that the random
experiment can in reality not be repeated for any given {\bf x} in situations
in which the researcher cannot control the {\bf x}.
Insisting on 
repeatability for the existence of traditional frequentist probabilities
would imply that nothing real corresponds to the distribution of $y$ for fixed 
{\bf x}, and there is no way to check any probability model. The unobservable 
i.i.d. \hbindex{error term} constructs 
\hbindex{repeatability} artificially, as in this way observations with different {\bf x} can be interpreted as involving repeated realisations of the 
error term. I have not seen this mentioned
anywhere. 
From a frequentism-as-model perspective this needs to be acknowledged,
and dependent on the situation 
it may be seen as appropriate or inappropriate, once more using information
from the data and about the subject matter background, 
but there is nothing essentially wrong or particularly suspicious about it, 
apart from the
fact that it turns out that our possibilities to test model assumptions are
quite generally 
more limited than many might think, see below.

\subsection{How is frequentism-as-model useful?} \label{suse}
From the discussion of \hbindex{mathematical models} and \hbindex{reality}
it follows that the separation between reality and 
model is a general feature of mathematical models, not an exclusive one
of frequentism-as-model, although frequentism-as-model acknowledges it 
explicitly.
The question ``how is frequentism-as-model useful, and how is it connected 
to reality?'' is
therefore connected with the general question how mathematical models
can be useful, even if
``wrong'' in Box's terms. The way to use mathematical 
models is obviously to take the mathematical objects that are involved in the
model as representations of either perceived or theoretically implied 
aspects of reality; and 
then to use this to interpret the results of using the models. 

In order for the model to be useful, does not reality have
to be ``somehow'' like the model? According to the constructivist view 
presented earlier, reality is accessible only through personal perception and
communication, so the best we can hope for is that the model can correspond to
how we perceive, communicate, and think about reality. Regarding 
frequentism-as-model, this concerns in the first place the concept of 
\hbindex{i.i.d.} \hbindex{repetition} as already discussed above. I.i.d. 
repetition is a thought 
construct and not an objective reality; however, even apart 
from probability modelling it is easy to see how much of our world-view even 
before any science relies
on the idea of repetition, be it the cycles of day and night and the seasons, 
or be it the expectation, out of experience, of roughly the same behaviour 
or even roughly the same distribution of behaviours when observing any kind
of recurring process as diverse as what a baby needs to do to attract the 
attention of the parents, look, size, and edibility of plants 
of the same 
species, or any specific  industrial production process. Objections against
the truth of ``i.i.d'' can be easily constructed for observations in 
all of these examples,
yet analysing them in terms of i.i.d. models often gives 
us insight, new ideas, and practical competence to deal with these processes,
be it with additional 
ingredients such as \hbindex{explanatory variables}, regular \hbindex{dependence} structures, 
or seasonal components.
Ultimately the success of probability modelling has to be 
evaluated empirically, relative to researchers' aims, and obviously the result  
will sometimes, but not always, be positive.
I remain agnostic regarding to what extent positive results confirm that 
``objective reality really is like that'', to which there ultimately is no 
answer beyond what occurs in personal and social reality. 

A major general reason for using \hbindex{probability models} 
is that they allow us 
to quantify \hbindex{random variation}. It is not an accident that probability
theory is a relatively young branch of mathematics. Unlike most  
historically older
mathematical modelling, 
probability modelling is
essentially not only about what was, is, or will be observed, but also about 
what could have been observed but was not and/or will not be observed. The 
latter does not and will not exist as observation. Connecting
this to observations is tricky and in my view a major reason why 
probability models are so controversial and problematic. But
they allow us to deal with a concept that was hardly acknowledged when 
probability theory emerged, and became hugely influential, 
namely the distinction between a \hbindex{meaningful pattern}
and \hbindex{meaningless variation}.

What can be done in order to support useful and avoid misleading
results of probability modelling
interpreted as frequentism-as-model? Below are some questions that need to
be addressed. These questions are
not very surprising, and statisticians, be they traditional frequentists, 
Bayesians, or something else, may think that they address these anyway. 
What makes frequentism-as-model different is that a standard approach would be
to take these issues into account when setting up and justifying the model,
and then all further analysis 
and interpretation is done conditionally on the model being true. In 
frequentism-as-model there is no such thing as a true model, meaning that the
issues below need to be kept in mind, and they still play a role 
when running model-based analyses and interpreting the results. 
\begin{description}
\item[(1) What is the aim of modelling?]
Probability models are used in different 
ways, and this is important for how to set them up. 
\cite{Hand94} emphasised the importance
of clarifying the \hbindex{research question} 
for statistical analysis, with several
examples illustrating how much of a difference this 
can make, regarding modelling and analysis, but also in earlier stages
such as the \hbindex{planning of experiments}.
Although the aim of 
modelling is not always ignored in traditional \hbindex{frequentist} or 
\hbindex{Bayesian} 
analysis, there is a general tendency to think that the ultimate aim is always
to find the true model, and if this (or at least a model as 
``close'' as possible to it) is found, everything ``relevant'' 
that can be known is known. 
This is fundamentally different from the attitude implied by 
frequentism-as-model. 

\cite{Cox90} lists three 
different classes of modelling aims, namely ``\hbindex{substantive models}'', which model 
(often causal) 
mechanisms that the researcher is interested in, ``\hbindex{empirical models}'',
which try to aid inference about relations between variables such as effect 
sizes, without making an attempt at reconstructing how the relations work in
any detail, and ``\hbindex{indirect models}'', where models are not directly connected 
with a real situation. 
This encompasses the use of models for deriving or
comparing methodology, but also, in a real situation but only indirectly 
serving  the main aim of analysis, things such as the imputation of missing 
values. The classes may be mixed or intermediate in some
specific situations.

This has methodological consequences. Substantive models may have aspects that
are there for representing the researcher's view but may refer to unobservables,
in which case they cannot or in the best case very
indirectly be tested against the data. A substantive model may even be there 
to communicate a researcher's view without the ambition to fit any data. It may
at times represent scenarios that are seen as best or worst cases, intentionally
not modelling what is seen as most realistic. \hbindex{Parsimony}
may be a strong concern if the model is used to communicate an idea, a weaker
concern if the model is used for prediction, and no concern at all if the model
is used to simulate data for exploring potential future variation in a complex 
system. In some cases the major use of the model is learning from discovering
its lack of fit and the exploration of reasons.

At first sight is seems central for
empirical models to fit the data. But also for such models the specific 
modelling aim is important, because deviations of the data from what would be 
typical for the model are only a problem to the extent that they raise the
danger of misleading results, which depends on what kind of result the
researcher is after, e.g., modelling integer number data by a Gaussian 
distribution is not normally a problem if only 
inference about means is required,
at least not as long as the i.i.d. assumption is implied and the 
distributional shape is not strongly skew or with large outliers, both of which
can happen with supposedly continuous data as well.

A major indirect use of models is the investigation of properties of 
\hbindex{inference} 
methods by theory or \hbindex{simulation}. 
For example, \hbindex{maximum likelihood (ML) estimators} 
can
be seen as inspired by probability models that are interpreted in a 
frequentism-as-model sense: ``Imagine a real situation as modelled, then 
the ML estimator makes the data most likely''. The major 
aim of modelling there is to find a good method for estimation. Even 
many Bayesians use models with an implicit frequentist interpretation in 
methodological work in order to investigate the quality of their methods in a
situation with a constructed known truth. It could be argued that in order 
for such
work to be applicable to objective reality, objective reality 
has to be as the model specifies, but if we accept that ``all models are 
wrong'' in objective reality, there is no better alternative than
studying the behaviour of method under a variety of such models. Simulations 
and theory will seem more relevant if the used models seem ``realistic'', but
as long as the aim is not to address a very specific application, models used
in this way are not checked against data. This only can happen when
the resulting methods are applied.

Furthermore, for some modelling aims there are methods that can more or less 
directly measure how well the aim is fulfilled. A major example is 
\hbindex{cross-validation} for \hbindex{prediction quality}. 
If prediction is the major 
modelling aim,
good cross-validation results can be seen as dominating 
other concerns such as fit of the 
existing data or correspondence with subject matter knowledge. 

Another example for direct measurement of what is of interest is the
objective function of \hbindex{$k$-means clustering} 
(for details see \cite{HeMeMuRo15}),
which 
directly measures the quality of
the approximation of any data point by the closest cluster centroid (more 
generally, often \hbindex{loss functions} can be constructed that formalise the
aim of analysis). If
achieving a good performance in this respect is the aim of analysis (for
example if clustering is used for data set size reduction before some other
analysis, and all clustered 
objects are replaced by the cluster centroid in a final analysis), $k$-means 
clustering is a suitable method even in situations that are very different from
the model for which $k$-means is ML, namely
Gaussian distributions
with same spherical covariance matrix in all clusters but different cluster 
means. Statisticians have argued that $k$-means often does not work well if
this assumption is violated (e.g., \cite{Vermunt11} in a discussion of 
\cite{SteBru11}), but this relies on the view that reconstructing the true model
is required, which may deviate from what is needed in
practice. 
\item[(2) Is the model compatible with the data?] I will discuss this in more 
detail in the subsection on statistical inference. 
\item[(3) How does the model relate to subject matter knowledge?] 
The model should be informed 
by perceptions and ideas (subject matter knowledge, which is not necessarily 
``objective'') 
regarding the real process to be modelled, such as specific 
reasons for dependence,
non-identity, or also for or against particular distributional shapes such as 
symmetry. Just to give an example, dependence by having the same teacher and 
communication between students makes the use of an i.i.d. model for analysing
student results for students from the same class or even school suspicious,
and it can be controversially discussed to what extent introducing dependence 
only in form of an additive random effect addresses this appropriately. 

\item[(4) How stable are conclusions against modelling differently?]  
The ultimate answer to the
question how we can be sure that a model is appropriate for the situation we
want to analyse is that we cannot be. \cite{Breiman01} mentioned the 
``multiplicity of good models'' in line with \cite{Tukey97,Davies14}, 
and
different ``good'' models may lead to contradicting conclusions. On top of that,
research hypotheses may be operationalised in different ways, using different
measurements, different data collection and so on. There is no way to be sure
of a scientific claim backed up by a statistical 
result based on the naive idea that the model is true and the method is
optimal. In order to arrive at reliable conclusions, science will need to
establish the \hbindex{stability} of the conclusions against different ways of 
operationalising the problem. This is basically the same way in which we as
individual human beings arrive at a stable concept of the world, as far 
as it concerns us; after lots of experiences, looking at something from 
different angles in different situations, using different senses, communicating
with others about it etc., we
start to rely on our ``understanding'' of something. In the same way, every 
single analysis only gives us a very restricted
view of a problem, and does not
suffice to secure stability, see the discussion of 
``\hbindex{interpretative equivalence}'' below. 
Currently there is talk of
a ``\hbindex{reproducibility crisis}'' 
in a number of disciplines (\cite{FidWil18}).
I think that a major reason for this is that no single analysis can establish 
stability, which is all too conveniently ignored, given that researchers and 
their funders like to tout big meaningful results with limited effort.
Even if researchers who try to reproduce other researchers' work have the
intention to take
the very same steps of analysis described in the original work that they try to
reproduce, more often than not there are subtle differences that were not
reported, like for example data dependent selection of a methodology that the
reproducer then uses unconditionally. Any single analysis will depend on 
researchers' decisions that are rarely fully documented, could often be replaced
by other decisions that are not in any way obviously worse, and may have a
strong impact on the result. Having something confirmed by an as large as 
possible number of analyses investigating the same research hypothesis of 
interest in different ways is the royal road to increase the reliability of
scientific results. In this I agree with the spirit of 
\cite{Mayo18}'s ``\hbindex{severity}'', and it also resonates with her  
``\hbindex{piecemeal testing}''; 
a scientific claim is the more reliable, the 
harder researchers have tried to falsify it and failed. Frequentism-as-model
surely does not license claims that anything substantial about the world 
can been proved by rejecting a single straw man null hypothesis. 
\end{description}
A further ``use'' of the consciousness and acknowledgement of the essential 
difference between model and reality may be a better awareness that probability
modelling is not mandatory; data can be often analysed descriptively or
visually without the need to compute p-values or posterior probabilities, and
in some situations \hbindex{quantification} itself can be regarded with more 
skepticism
\cite{TaSlNe16,Chang22}.

\subsection{Frequentism-as-model for statistical inference} \label{sinference}
Adopting frequentism-as-model as interpretation of probability does not imply 
that classical methods of frequentist inference, \hbindex{tests}, 
\hbindex{confidence intervals}, or
\hbindex{estimators} 
have to be used. It may also be combined with Bayesian inference as discussed later. However, the classical methods of \hbindex{statistical 
inference} have valid frequentism-as-model interpretations, and I believe that 
understanding these interpretations properly will help to apply the methods in
a meaningful and useful way. Therefore I disagree with recent 
calls to abandon
\hbindex{significance testing} 
or even frequentist inference as a whole (\cite{WSL19}). 
\subsubsection{Interpretative equivalence, stability, and robustness}
\label{sinter}
The question how stable our conclusions are as raised above
does not only depend on models,
methods, and the data, but also on the conclusions that are drawn from them, 
i.e., the interpretation of the results by the researcher. A helpful concept
to clarify things is ``\hbindex{interpretative equivalence}''. 
I call two models 
``interpretatively equivalent'' with respect to a researcher's aim if
their subject-matter interpretation of what they are interested in is the same
regardless of which of the two models is true. For example, if a researcher
is interested in testing whether a certain treatment improves blood pressure 
based on paired data before and after treatment, she will in all
likelihood draw the same conclusion, namely that the treatment overall
does not change the blood pressure, 
if the distribution of differences between after and before treatment 
is symmetric about zero, be it a Gaussian or a $t_2$-distribution, for example.
For her research aim, the precise shape of the distribution is irrelevant.
Comparing two models with expectation zero, one of which is symmetric whereas
the other one is not, this is not so clear; the treatment changes the shape of 
the distribution, and whether this can be seen as an improvement may depend 
on the precise nature of the change. For example, consider the model
\begin{equation}\label{eoutlier}
0.99{\cal N}(\mu^*,\sigma^2)+0.01*\delta_{100},
\end{equation} 
$\delta_{x}$ being the one-point
distribution in $x$. 
The researcher may think of  $\delta_{100}$-distribution 
as modelling erroneous observations, in which case $\mu^*=0$ makes the
model interpretatively equivalent to ${\cal N}(0,\sigma^2)$, whereas 
for $\mu^*=-\frac{1}{0.99}$, despite expectation zero, the researcher
will interpret an average effect of lowering the blood pressure, which the
model implies for 
a subpopulation of 99\% of people.

In fact, the term ``interpretative equivalence'' implies an oversimplification,
as models can be more or less interpretatively similar, and it is not always
possible or helpful to make a binary distinction between ``interpretative 
equivalence'' and ``interpretative difference'', but I will stick to this
distinction here for ease of discussion. 

The importance of interpretative equivalence is that it allows to discuss what
deviations of a temporarily assumed model should be handled in which way. If 
conclusions are drawn from a method involving certain 
\hbindex{model assumptions}, it 
would be desirable, and conclusions can be seen as stable, if the probability 
would be roughly the same that the researcher arrives at the same conclusion 
from the same data if data
are actually generated by an interpretatively equivalent model other than the
assumed one. If this is not the case, it would be
of interest if the models can be told apart with good probability by the data,
in which case the data can be used to decide on which one of these models to
base a data analysis. This also applies to the decision between a 
specific \hbindex{parametric} and a more general but potentially less precise
\hbindex{nonparametric} approach. 

The concept also clarifies the issue that tests on \hbindex{large samples} 
reject point \hbindex{null hypotheses} too easily. 
Considering a $\mu=0$ null hypothesis, a model with 
$\mu\neq 0$ but a very small absolute value will lead
to a rejection of $\mu=0$ with a large probability for large $n$. This is a 
problem if and only if models with, say, $|\mu|\le\epsilon,\ \epsilon>0$ are
considered interpretatively equivalent to the model with $\mu=0$, in other
words, if a difference as small as $\epsilon$ is considered  
irrelevant from a subject matter point of view. 
If this can be specified, the question whether values of $\mu$
with $|\mu|\le\epsilon$ are in the corresponding \hbindex{confidence interval}, 
or what
\hbindex{severity} 
the test achieves ruling out $|\mu|\le\epsilon$, see \cite{Mayo18},
is more relevant than whether a test of $\mu=0$ is significant. 
Note that in a real situation it will
hardly be possible to specify a precise
borderline $\epsilon$ and one could think that things are rather continuous 
than binary, however due to the 
discreteness of language, in many situations at least certain nonzero 
differences will be called irrelevant.

Being concerned about interpretative equivalence of models is an entry point
for \hbindex{robust methods}, which are designed for dealing with the possibility 
that a nominal model is violated in ways that are hard to diagnose 
but may make
a difference regarding analysis. \hbindex{Robustness theory} (e.g., 
\cite{HRRS86,HubRon09}) is about limiting 
changes in results under small changes of the modelled processes. An issue with
standard robustness theory is that it is often
implied that ``contamination'' of
a distribution should not have an influence of the results, whereas in practice
the contamination may actually be meaningful. To decide this is a matter
of assessing interpretative equivalence. Frequentism-as-model 
acknowledges that desirable behaviour of a statistical method is not just 
something ``objective'' but depends on how we interpret and assign meaning
to differences between distributions. Robustness is helpful where the reduced
sensitivity of robust methods to certain changes in the data or model takes
effect where the changes are indeed interpreted as meaningless relative to
the aim of analysis; for example there needs to be a decision whether in model
(\ref{eoutlier}) what correspond to the quantity of interest is rather the
mean of the distribution, or rather the $\mu^*$.  
In some applications it is inappropriate to use a method 
of which the results may not change much when replacing up to half of the data,
namely where the resulting processes are considered as interpretatively very 
different,
but in any case comparing the results of such a method with a more
``sensitive'' non-robust one may allow for a more differentiated perception 
of what is going on than does every single method on its own.  

\subsubsection{Significance tests, frequentist inference} \label{stest}
Statistical \hbindex{hypothesis tests} have become very controversial 
(e.g., \cite{WSL19}), but the interpretation of tests from the position of 
frequentism-as-model is rather straightforward. They address to what extent 
models
are \hbindex{compatible} with the data in the sense formalised by the test statistic.
For simplicity, I mostly call models ``compatible'' or ``incompatible'' 
with data, but in fact compatibility is gradual, as expressed 
by \hbindex{p-values}.

Obviously a non-rejection cannot be an indication that the model is in fact 
true, but it means the absence of evidence in the data 
that reality is different in the way implied by the test statistic,  
making it impossible to claim such evidence. 
Sometimes tests, particularly two-sided tests, are criticised because 
``point null hypotheses are usually scientifically implausible 
and hence only a straw man'' (e.g., \cite{RBK20}), but then a parametric model
allowing for any parameter value cannot be ``really true'' anyway, so that
for the same reason for which the point \hbindex{null hypothesis} ($H_0$) 
is ``implausible'',
the whole parametric family is implausible as well, but that does not 
necessarily stop
inference about it from being informative and useful. 
What a  
test can do is neither to confirm $H_0$ as true, nor to allow to
infer any specific alternative in case of rejection. A two-sided test should
be run if compatibility of the data with $H_0$ is a  
possibility of interest; even then it may not be the method of choice if the 
sample size is too large, see the earlier discussion. Recall that tests are 
often chosen according to \hbindex{Neyman-Pearson logic} \citep{Lehmann86}.
This means that they are meant to maximise the \hbindex{power}, i.e., to
minimise the probability for non-rejection
given a specific alternative hypothesis, and given a fixed value for the 
probability to reject a true $H_0$. This is often read as being 
only meaningful if either the nominal $H_0$ or the specific alternative are 
true. Instead, I see this reasoning as a rather clever idea to motivate test 
construction in a real situation, using mathematical modelling. A reasonable 
formal criterion is optimised in the model-world, leading to a procedure that
can be applied in the real world, which differs from the model-world. It is 
important to ask to what
extent the differences between the real world and the model-world have the 
potential to mislead the procedure, but this can only be assessed case by case.
For sure it cannot be required that the real situation satisfies the formal 
model-optimality. Demanding this would be a category error.

Often significance tests are presented implying that a \hbindex{rejection} of the $H_0$
is meaningful whereas a non-rejection is not. Above I have explained the
meaning of a non-rejection in frequentism-as-model; the meaning of 
rejection is a more complex issue. Obviously, rejection of the $H_0$ 
does not imply that
there is evidence in favour of any particular model that is not part of the
$H_0$. On the positive side, a rejection of the $H_0$ gives information about
the ``direction'' of deviation from the $H_0$, which can and mostly should 
be substantiated using \hbindex{confidence sets} 
of compatible parameter 
values, and particularly data visualisation for exploring how this plays out
without relying on the specified model. 

Consider a \hbindex{one-sample t-test} of the $H_0:\ \mu=0$ 
regarding ${\cal N}(\mu,\sigma^2)$  
against $\mu>0$. The test
statistic $T$ 
is the difference between the sample mean and $\mu=0$ 
standardised by 
the sample standard deviation. 
Sticking to model-based thinking while not 
implying that ${\cal N}(\mu,\sigma^2)$ is true, 
rejection of the $H_0$ can be interpreted as providing
evidence in favour of any distribution of the two samples
for which a larger value of $T$ is
observed with larger probability than under $H_0$, compared with the $H_0$.
We do not only have evidence against
$H_0$; we also learn that the problem with $H_0$ is that $\mu=0$ 
is most likely too low, which is informative.    

In many situations in which the
one-sample t-test is used, the user is interested in testing $\mu=0$ or rather
its interpretative meaning, 
but not
in the distributional shape ${\cal N}(\mu,\sigma^2)$, which just enables them
to run a test of $\mu=0$. If in fact there were a true distribution, which of
course we can consider, in the ``as if''-model world, which isn't 
${\cal N}(\mu,\sigma^2)$ with any specific $\mu$ or $\sigma^2$, it is not clear
in a straightforward manner what would correspond to $\mu$. It may be the 
expected value, but in a model such as (\ref{eoutlier}) it may well be seen 
as the $\mu^*$ that governs 99\% of the observations. In any case, there are
distributions that could be considered as \hbindex{interpretatively equivalent}
to the actual $H_0$. Any 
distribution symmetric about zero is in most applications of t-tests 
interpretatively equivalent to the $H_0$, meaning that if indeed a 
distribution different from ${\cal N}(\mu,\sigma^2)$ but still symmetric about
$\mu=0$ were true, it would be desirable that the probability to reject 
$H_0$ would be as low as if $H_0$ were true. 

One can then ask how likely it is 
to reject the $H_0$ under a model that is
not formally part of the $H_0$, but interpretatively equivalent.
For the one-sample t-test, assuming
existing variances, many distributions interpretatively equivalent 
to ${\cal N}(\mu,\sigma^2)$ are asymptotically equivalent as well, and 
results in \cite{Cressie80} 
and some own simulations indicate that it is very hard if not impossible even
in finite samples to generate an error probability of more than 7\% 
for rejecting an interpretatively true $H_0$ with a nominal level of 
5\%, even if the underlying 
distribution is skew and $\mu$ is taken to be the expected value, meaning that
testing at a slightly smaller nominal level than the maximum error probability
that we want to achieve will still allow for the intended interpretation. 

But the situation
becomes much worse under some other deviations from the \hbindex{model assumptions}, 
particularly positive dependence between observations, which increases the
variation of $T$ and may lead to significant $T$ and increased type I
error probability very easily, also in situations that are 
interpretatively equivalent to the $H_0$. Consider an example in which 
the expected change of the turnover generated 
by a salesperson after attending a sales seminar is of interest in order to 
evaluate the quality of the instructor.
If all 
salespersons in the sample attend the sales seminar together, they may learn 
something useful from talking with each other about sales strategies, even if
what the instructor does itself is useless.   
 
What is required in order to achieve a reliable conclusion is just what was
discussed in the section on the use of models. The general question to 
address is whether
a rejection of $H_0$ could have been caused by something interpretatively
equivalent to the $H_0$, in which case the conclusion could not be relied upon.
The questions to ask are: (1) is there any evidence in the data that such a 
thing may have happened, (2) is there any subject matter knowledge that suggests
this (e.g., salespersons may have learnt from each other rather than from the 
seminar), (3) does the meaning of the test statistic correspond to what is of
interest, and (4) is it feasible to run further analyses, with the same or 
other data, that can confirm the conclusion? (4) is important because even the
best efforts to use (1) and (2) will not be able to remove all doubt. Models
are
always conceivable that can cause trouble but cannot be distinguished from 
the nominal model based on the data, and thinking about the subject matter may 
miss something important. 
(3) is in my view a very important issue that is
not usually appreciated (but see \cite{Hand94}). 
Tests are usually derived using optimality 
considerations under the nominal model, but this does not imply that they are
optimal for distinguishing the ``interpretative $H_0$'' from the 
``interpretative alternative''. In many cases at least they make some good 
sense, but a model like (\ref{eoutlier}) may lead them astray. 
The full interpretative $H_0$ and the 
interpretative alternative will normally be too complex to derive any
optimal test from them, but the form of the \hbindex{test statistic} 
itself suggests
what kinds of distributions the test actually distinguishes. This should be in
line with what is interpreted as the difference of interest between $H_0$ and
the alternative. At least under a point $H_0$, the distribution of any 
test statistic can be simulated (if the $H_0$ is a set of distributions, 
parametric bootstrap can be used, if potentially involving a certain bias).
Assuming that rejection happens for large values of the test statistic,
it implicitly defines the  interpretative $H_0$ and the 
interpretative alternative distinguished by the test as the set of 
distributions for which the test statistic is expected low,
 and the set of distributions for which it
is expected larger, respectively. 
In many situations a nonstandard test statistic may be 
better suited than what is optimal under a simple nominal model (see 
\cite{HenLin15} for an example).

Bayesians often argue against frequentist inference by stating that the 
probabilities that characterise the performance of frequentist inference methods
such as \hbindex{test error} and \hbindex{confidence set} 
\hbindex{coverage probabilities} are 
pre-data probabilities, and that they do not tell the researcher about the 
probability for their inferences to be true after having seen the data. 
One example is that a 95\%-confidence interval for the mean of a Gaussian 
distribution in a situation in which the mean is constrained to be larger 
than zero may consist of negative values only, meaning that after having seen 
the data the researcher can be sure that the true value is not in the 
confidence interval. This is a problem if the coverage probability of the 
confidence interval is indeed interpreted as a probability for the true 
parameter value to be in the confidence interval, which is a misinterpretation;
the coverage probability is a \hbindex{performance characteristic} of the 
confidence interval pre-data, assuming the model. From the point of view of 
frequentism-as-model,
this is not a big problem, because there is no such thing as a true model or a 
true parameter value, and therefore a probability for any model to be true is
misleading anyway, although it can be given a meaning within a model in which 
the prior distribution is also interpreted in a frequentism-as-model sense,
see below. In practice, if in fact a confidence interval
is found that does not contain any admissible value, this means that either
the restriction of the parameter space that makes all the values in the 
confidence interval impossible can be questioned, or that a truly atypical 
sample was observed. \cite{Davies95} defined ``\hbindex{adequacy regions}'' that
are basically confidence sets defined based on several statistics together
(such as the mean or median, an extreme value index, and a discrepancy between 
distributional shapes; appropriately adjusting confidence levels), 
that can by definition in principle rule out all distributions
of an ``assumed'' parametric family, meaning that no member of the family is
compatible with the data given the combination of chosen statistics. 
Normally confidence intervals are interpreted as giving a set of truth 
candidate values assuming that the parametric model is true; but if that 
assumption is dropped from the interpretation and it is just a set of models
compatible with the data defined by parameter
values within a certain parametric family, it is possible that the whole family 
is not compatible with the data. 

Overall I suspect that the biggest problem with interpreting standard 
frequentist inference is that many of the people who use it want to make
stronger claims, and want to have bigger certainty, than the probability
setup that they use allows. An example for this is the ubiquity of implicit
or explicit claims
that the \hbindex{null hypothesis} is true in case that it was not rejected by a test. 
Interpreting the results of classical frequentist inference assuming the
truth of the model will generally lead to overinterpretation. According to
frequentism-as-model, classical 
frequentist inference allows compatibility and incompatibility statements of
models with data. As far as the models are used to make substantive statements,
such compatibility assessment is of course of interest.
What exactly can be learnt from this depends on the 
analysed situation as well as on what the researcher sees as 
interpretatively equivalent in that situation. From the point of view of
frequentism-as-model, research about the expected variation of results under 
interpretatively equivalent models would be desirable, acknowledging that
interpretative equivalence may mean different things in different 
applications. Deciding on a parametric model and from then on ignoring that 
other models can be compatible with the data as well will easily lead to 
overinterpretation.

\subsubsection{Model assumptions}\label{smodelass}

A standard statement regarding statistical methods is that these are based
on \hbindex{model assumptions}, and that the model assumptions have to be 
fulfilled 
for a method to be applied. This is misleading. Model assumptions cannot be
fulfilled in reality, because models are thought constructs and operate on
a domain different from observer-independent reality. 
For this reason they cannot even be
\hbindex{approximately true}, 
in a well defined sense, because no distance between
the unformalised ``underlying real truth'' and the model can be defined, 
although, following Davies, the approximation notion can 
be well defined comparing observed data to a model.

What is the role of model assumptions then? There are theorems that grant a 
certain performance of the method, sometimes optimal, sometimes just good in
a certain sense, under the model assumptions. The model assumptions are not
required for applying the method, but for securing the performance achieved in
theory. The theory can be helpful to choose a method, and some
theory leads to the development of methods, but
the theoretical performance can never be 
granted in reality. This does not mean that the performance will be bad 
whenever the model assumptions are not fulfilled. In 
fact, some aspects of the model assumptions are usually almost totally
irrelevant for the
performance, such as the application of methods derived from models for
continuous data to data that is rounded to, say, two decimal places. 
\hbindex{Interpretative equivalence} 
is an important concept also in this respect,
because optimally methods would not only distinguish what is deemed 
relevant to distinguish 
by the researcher 
under the model assumptions, e.g., a certain $H_0$ and its nominal alternative,
but they would lead to largely the
same results for interpretatively equivalent distributions that do not fulfill
the model assumptions. As far as this is the case, the corresponding model
assumption is irrelevant.

It makes sense, at first sight, to think that the method will be appropriate
if the assumed model is a good model, i.e., if reality looks very much like it. 
This itself can be modelled, meaning that one can look at the performance of
a method in a situation in which the assumed method does not hold, but a
somewhat similar model, which can be formalised using dissimilarity measures
between distributions. This has been considered in \hbindex{robust statistics} 
(\cite{HRRS86,HubRon09}, ``qualitative robustness'' in particular) with the 
unsettling
result that the performance of some classical statistical methods can decline
strongly in arbitrarily small neighbourhoods of the assumed model. As a 
consequence, more robust methods have been developed, 
i.e., methods with a more stable 
performance in close neighbourhoods. This is surely valuable progress, but 
on one hand it does not change the fact that models can be set up that cannot
be rejected by any amount of data, and that can annihilate the performance of
any method, e.g., using irregular dependence and/or non-identity structures, so
full safety cannot even be had in the modelled world, let alone the real one. 
On the other hand, the 
situation is not quite as hopeless as this may suggest. Usually bad results 
about standard methods are worst case results, very often regarding extreme 
outliers, and it is often plausible, from checks, \hbindex{data visualisation}, 
or subject
matter considerations, that the worst case does not occur. To argue against
detrimental dependence or non-identity structures is much harder, and the best
that can be done in this respect is looking for conceivable reasons for 
trouble, using a suitable method if no such reason comes to mind, and then 
hope that the world will behave when acting based on the result. Science makes
mistakes and is hopefully open-minded enough to correct itself in case of bad 
outcomes.

\cite[Sec. 4.8-4.11]{Mayo18} and others argue that model assumptions can 
and should be 
tested. Whereas the truth of the model cannot be secured, much can indeed 
be found
out about whether a model is \hbindex{adequate} for a purpose. But 
\hbindex{model checking} has a number of issues. The first one is that 
if a certain model
assumption turns out to be compatible with the data, it does not mean that 
other models that may lead to different results are ruled out. 
\hbindex{Misspecification testing} cannot secure robustness
of results against models that cannot be excluded.

A second issue is that if a method is used
conditionally on passing a model misspecification test for checking a model
assumption, the distribution of the data that eventually 
go into the method becomes 
conditional on passing, and this will normally violate the model assumptions, 
even if they were not violated before. Particularly it will make observations 
dependent even if they were not dependent before, because if for example $n-1$
observations are on the borderline for passing the misspecification test, the
$n$th observation has to fit well for passing. I called this ``goodness-of-fit
or \hbindex{misspecification paradox}'' in \cite{Hennig07}. In many 
situations (but not always) this
will not change error probabilities associated with the model-based method
strongly, meaning that in case the model was true before misspecification
testing, not much harm is done.

On the other hand, in case that the model assumptions are violated in
a problematic way, the distribution conditionally under passing a 
misspecification test will often not make the method work better, and 
sometimes even 
worse than before testing; keep in mind that just because the misspecification
test does not reject the model assumption, it does not mean that it is 
fulfilled. \cite{ShaHen19} reviewed work investigating the
actual performance of procedures that involve a misspecification test for one
or more model assumptions before running a method that is based on these
model assumptions, looking at data that originally fulfilled or did not fulfill
the model assumption. Occasionally a \hbindex{combined procedure} 
was investigated in
which, in case of the violation of a model assumption, a method with lighter 
assumptions is used. The philosophy of such work, which is in line with
frequentism-as-model, is that the key question is 
not whether the model assumptions are really fulfilled, but what the performance
of such combined procedures is in several situations, compared to both the 
model-based and the less model-based (often \hbindex{nonparametric}) test. 
The results
are mixed, depend on the specific combinations of tests, and 
surprisingly many authors advise against misspecification testing
based on their results, e.g., \cite{FayPro10}: 
\begin{quotation}
The choice between t- and Wilcoxon-Mann-Whitney
decision rules should not be based on a test of normality.
\end{quotation}  
Whereas formally requiring lighter model 
assumptions, the use of \hbindex{nonparametric methods} 
only pays off if the probability to arrive at a conclusion
that is interpretatively in line with a modelled truth is better than for
a competing parametric method for most models compatible with data and existing
knowledge; traded off potentially against the expected sizes 
of ``interpretative 
differences'', as far as these can be specified. This is not always the case.

\cite{ShaHen19} showed a theoretical result that presents combined procedures
in a somewhat more positive light. They looked at the overall performance of
combined methods in a setup where datasets could be, with a certain probability
$\lambda$, generated by the assumed model, and with probability $1-\lambda$ by 
a distribution that could cause trouble for the model-based method
(this probability could be taken as rather idealised aleatory one considering
many studies of a similar kind).
Under certain assumptions for the involved methods they
showed that for a range of values of $\lambda$ the combined procedure beats both
involved tests, the model-based one and the one not requiring the specific 
model assumption, regarding
power, even if
not winning for $\lambda=0$ and $\lambda=1$, which is what authors of previous 
work had investigated.

The surprisingly pessimistic assessment of misspecification testing by authors
who investigated its effect may be due to the fact that most available 
misspecification tests test the model assumption against 
alternatives that are 
either easy to handle or very general, whereas little effort has been spent on
developing tests that rule out specific violations of the model assumptions
that are known to affect the performance of the model-based method strongly,
i.e., leading to different conclusions for \hbindex{interpretatively 
equivalent} models,
or same conclusions for models that are interpretatively very different.
Such tests would be specifically connected to the model-based method with which
they are meant to be combined, and to the assessment of interpretative 
equivalence. 

Furthermore, as already mentioned, 
not everything can be tested on data. 
E.g., many conceivable 
\hbindex{dependence} structures 
do not lead to patterns that can be used to reject 
independence. For example, it may 
occasionally but not regularly 
happen that one observation determines or changes the distribution
of the next one.
Think of psychological tests in which sometimes a test 
person discusses the test with another participant who has not yet been tested,
and where such communication can have a strong influence on the result.

Frequentism-as-model could inspire work that looks at performance of 
methods involving model assumptions and
misspecification testing under all kinds of models that seem realistic,
including Bayesian 
combinations of different models with different probabilities. 
Generally, 
regarding methodological research, a researcher adhering to  
frequentism-as-model will always be interested in the performance of methods 
that are supposedly model-based under situations in which the model assumptions
are violated, knowing that model assumptions can never be relied upon, and
that looking at other models is the only way to investigate what happens 
then. This is in the tradition of \hbindex{robust statistics}, but with less 
focus
on worst cases and more focus on interpretative equivalence for deciding what
outcome would actually be desirable. 
The cases that are really the worst ones are hopeless under any
approach, and even the best statistics cannot always save the day.

\subsubsection{Frequentism-as-model and falsificationist Bayes} \label{sfambayes}

\hbindex{Epistemic probabilities} 
model a personal or ``objective'' \hbindex{degree of belief}, 
not the \hbindex{data generating process} as such, and therefore they cannot 
be checked against and falsified by the data. See \cite{Dawid82} for a 
discussion of ``\hbindex{calibration}'', i.e., agreement or potential mismatch between 
predictions based on epistemic Bayesian probabilities and what is actually 
observed.   
\cite{GelSha13} argued that \hbindex{Bayesian statistics} 
should allow for checking the 
model against the data, and interpret Bayesian models as modelling data 
generating processes rather than epistemic uncertainty. This is in line with
a lot of applied Bayesian work in which the posterior distribution of the 
parameter is interpreted as encoding probabilities for a certain parameters 
being true descriptors of the underlying data generating process. 
This
allows to check the model against the data and potentially to revise it, even
though this violates the coherence requirement, which is taken as essential
by most advocates of epistemic Bayesian interpretations.
\cite{GelHen17} called the connection of inverse probability logic and 
model checks ``\hbindex{falsificationist Bayes}''. 

My interpretation of it
is that probabilities in falsificationist Bayes are interpreted in the same 
manner as in frequentism-as-model. The parametric model is handled {\it as if}
it describes the data generating process but can be dropped or modified
if falsified by the 
data. The involved interpretation of probabilities is consistently aleatory 
if the
\hbindex{parameter prior} is interpreted as a model of a parameter generating process
in the same way. \cite{GelSha13} stated that the 
\hbindex{prior distribution} may encode ``{\it a priori} knowledge'' or a 
``subjective degree of belief''. This seems to mix up an 
\hbindex{epistemic interpretation} of the parameter prior with an 
\hbindex{aleatory interpretation} of the parametric model, and it is hard to 
justify using them in the same calculus. I believe that it would be better 
to refer to the parameter prior as an idealistic model of a process that 
generates parameters for different situations that are based on the same 
information, i.e., to interpret it in a frequentism-as-model way. This is in 
line with the fact that Gelman in presentations sometimes informally
refers to the parameter prior as a distribution over parameters realistic in a
distribution of different situations of similar kind in which datasets can be 
drawn. This is a very
idealistic concept and it is probably hard to connect the setup of parameter
generation precisely to real observations. The modelling will normally indeed
rely more on belief and informal knowledge than on observation of replicates of
what is supposed to be parameter generation, 
but an arbitrary 
amount of data can be
generated from the fully specified model, and can be compared with the 
observed data. Testing the parameter prior is hard. In a standard simple
Bayesian setup, it is
assumed that only one (possibly multidimensional) 
parameter value generated all observed data, so
the effective sample size for checking the parameter prior is smaller than one,
because the single parameter is not even precisely observed. Therefore 
sensitivity 
against prior specification will always be a concern, but see the earlier
discussion
of single-case probabilities. In any case, the parametric model can be 
tested in frequentist ways. In case the parameter prior
encodes valuable information about the parameter that can most suitably be
encoded
in this way, Bayesian reasoning based on such a model is clearly useful, and 
open to self-correction by falsificationist \hbindex{compatibility logic}. 
As for model assumptions
testing followed by traditional frequentism methods (\cite{ShaHen19}), 
it may be of interest to
analyse the behaviour of Bayesian reasoning conditionally on model checking
in case of fulfilled and not fulfilled model assumptions.

A benefit of such an interpretation could be that 
the prior no longer either has to be claimed to be objective or to model
a specific person. It is a not necessarily unique 
researcher's suggestion how to imagine the parameter
generating process based on a certain amount of information, and can as such 
be compared with alternatives and potentially rejected, if not by the data, then
by open discussion. As earlier, \hbindex{stability} 
against different prior choices 
compatible with the same information can be investigated. Models may be set up
not only to formalise most realistic processes, but also, depending on the 
application, worst or best case scenarios in order to explore what range of
possibilities this generates. 

Alternatively, the sampling model may be interpreted in a frequentist manner, 
and the parameter prior may not be of interest in its own right, but rather
be used as a tool to achieve inference with good
\hbindex{frequentist properties} such as \hbindex{confidence coverage} or
small \hbindex{estimator} \hbindex{mean squared error}
\citep{BayBer04,Wasserman06}.
In this case, the posterior probabilities cannot be 
interpreted, as the posterior inherits meaning from the prior, and the prior
has not been chosen in order to be meaningful. Often, Bayesian methods have 
better frequentist properties for certain parameter values and worse for 
others, so that the prior can also be used to improve matters for certain 
parameter values that are seen as more relevant, to the price of having a worse
result in some other cases, being potentially in line with the idea that
\hbindex{prior information} 
suggests that certain parameter values are more likely 
than others.


Falsificationist Bayes combines the \hbindex{inverse probability logic} and 
compatibility logic
by applying the former
within a Bayesian model, of which the compatibility with the data should also
be investigated. \cite{GMS96} emphasise that the posterior distribution of the 
parameter is conditional on the truth of the model, and according to
frequentism-as-model a researcher can interpret results temporarily {\it as if}
this were the case, without ignoring that this is just a thought construct and
that many other models are compatible as well. 
This differs from the epistemic probability interpretation, where the
``truth of the model'' is not a matter of mechanisms in the real world, but 
rather of the degrees of belief of the analyst,
and the possibility that the analyst's degrees of belief are not properly
reflected by the model is rarely discussed.

Frequentism-as-model allows to interpret the parameter prior in falsificationist
Bayes in a way
that requires neither to mix epistemic and frequentist meanings of probability,
nor to demand that the prior corresponds to 
an in principle infinitely repeatable data generating process. 
The Achilles heel of this is that in a standard situation, based 
on a single realisation that is not even directly observable, the potential
to check the parameter prior against the data is very weak. The prior still
needs to be convincingly defended  in other ways, and sensitivity analysis is
certainly desirable.

\section{Epistemic-probability-as-model}\label{sepam}
As for aleatory probability, it is also possible to use \hbindex{epistemic probability}
in a way that explicitly takes into account that models are tools for thought
rather than referring to any truth, and be it a truth concerning a subjective
agent.

This chapter had the main focus on frequentism-as-model, and I will only make 
short remarks on \hbindex{epistemic-probability-as-model}. 
Some of the discussion 
regarding frequentism-as-model also applies to epistemic-probability-as-model,
such as regarding the role of the \hbindex{exchangeability} 
assumption, which is analogous
to the role of \hbindex{i.i.d.} in frequentism as discussed earlier. 
Considerations
regarding \hbindex{interpretative equivalence} and \hbindex{robustness}
are also relevant. 

\subsection{Epistemic idealisation}

As the idea of limiting relative frequencies, 
the \hbindex{axioms} of epistemic probability are \hbindex{idealisations}, and as such 
epistemic probability models are, as models, essentially different from
reality, which here is the reality of agents reasoning under uncertainty. 
Adopting such a model amounts to adopting an idealisation, and the ``as-model''
approach taken here states that this should not be forgotten. In fact,
idealisation is conceded in many places (e.g., \cite{HowUrb06}), but as in
frequentist analyses, results in practice are often interpreted as if 
idealisation has not happened (for example probabilities that
a certain parameter lies in a suitably chosen set may be portrayed without
stating relative to which choices this has been computed, or without much
justification of these choices). 

De Finetti's 
(and others') arguments regarding coherent \hbindex{betting}  
require linear utility of money. Leaving this issue aside, the involved
betting regarding completely specified distributions is hardly ever done 
in practice; if at all (as for example in some financial applications), 
human agents with specific points of view bet against other human agents,
and taking into account knowledge about the opponent or more generally human 
behaviour may influence the offered bets. Several authors 
\citep{Jaynes03,HowUrb06} prefer the
derivation of the probability axioms from desiderata for rational reasoning 
about plausibility under uncertainty as in \cite{Cox46}'s 
Theorem to the betting approach, but also this involves idealisation (i.e.,
the assumption that such plausibilities always can be expressed by bounded real 
numbers; any two events can be compared on a one-dimensional scale, and a 
number of other not necessarily intuitive requirements, see \cite{VanHorn03}), 
and it 
does not give an operational meaning to the probabilities chosen in any specific
situation. 

There are two levels at which idealisation takes place. The first one 
is the choice of \hbindex{priors}. It has already been mentioned 
that prior choices in practice will necessarily be simplified
by not taking everything into account that could possibly happen. Furthermore,
there are different principles for designing \hbindex{informationless priors} 
that in
a given situation may lead to different choices and different results 
\citep{KaWa96}. The second level is the justification of Bayesian (or other)
updating as a supposedly objective way to process probability statements given
new data, whether or not the priors are taken as objective. But this second
level can hardly be considered independently of the prior choice, as modifying
an ill-chosen prior may require a deviation from Bayesian updating and 
coherence. See \cite{VanHorn03} for a comprehensive
discussion of other possible objections to the desiderata of Cox's Theorem.  

\subsection{Flexible handling of priors}\label{sprior}
Many pragmatic Bayesians have been well aware for a long time of the 
problems of all too dogmatically following the axioms of epistemic probability,
and many recommendations exist that deviate from these, so that many practical 
implications of epistemic-probability-as-model are not new. E.g., 
\cite{BerSmi00} wrote:
\begin{quotation}
In this work, we shall proceed on the basis of a
prescriptive theory which assumes precise quantification, but then pragmatically
acknowledges that, in practice, all this should be taken with a large pinch 
of salt and
a great deal of systematic sensitivity analysis.
\end{quotation}
More 
suggestions and
references can be found in \cite{GCSDVR13}, particularly in chapter 6. 

Dropping the requirement to find a ``true'' prior, it is possible to explore
a range of priors. For example, \cite{SpFrPa94} propose to look at a 
``skeptical'',
an informationless, and an ``enthusiastic'' prior in randomised trials, in order
to see clearly how posterior results are affected by different attitudes 
expressed in the prior. \hbindex{Sensitivity analysis} 
looking at different priors that 
are seen as encoding (more or less) the same prior information is sometimes
used. It is also possible to choose a prior that intentionally only uses
certain prior information (that may be uncontroversial) while not accounting for
other information that may be controversial among stakeholders, to allow for 
better agreement. See \cite{Dawid82a} on intersubjective choices of epistemic
probabilities.
 
Maybe the biggest problem with the \hbindex{coherence} requirement for 
epistemic 
probabilities is that it does not allow the user to deviate from the prior
choices other than by \hbindex{Bayesian updating}, see above. 
Pragmatic Bayesians know that
it is practically impossible to specify prior probabilities for everything
that can possibly happen. In case of a conflict between prior and data
(to be discovered using \hbindex{compatibility logic}, temporarily), 
it is often seen as more appropriate to change the model, see, e.g., 
\cite[Ch. 6]{GCSDVR13}, even if this
violates the \hbindex{axioms} of epistemic probability. 
According to the \hbindex{betting}
logic proposed by Ramsey and de Finetti, this will allow an opponent to choose
bets that incur sure loss for the user who chooses to change their model.
But of course the betting setup is itself an idealisation, and in much applied
data analysis, bets based on epistemic probabilities do not really happen.
It may therefore seem harmless to ``\hbindex{reject}'' earlier prior choices.  
\cite[Ch. 6]{BerSmi00} back this up by utility considerations (incurring sure
loss may be milder than exposing oneself to higher losses due to an 
inappropriate model choice), but they 
admit that
rejecting a model cannot be fit into the Bayesian framework unless alternative
models are already specified. Another reason to change priors is \hbindex{miscalibration},
i.e., bad prediction results \citep{Dawid82}. 

It should not be ignored that if the data analyst is willing to 
violate the axioms for epistemic probability, these axioms can no longer be
used to sell the Bayesian approach as \hbindex{coherent} 
and unified. All Bayesian 
posterior probabilities are functions of the \hbindex{prior}. If the prior is
changed conditionally on the data, the interpretation of the posterior is 
questionable. A way to still justify the resulting \hbindex{posterior} 
probabilities 
is to argue that a prior may have been specified in advance 
that involves both the original
and the changed model, and given the incompatible 
data the posterior probability for the original model should be very small, 
so that
posterior probabilities from the new changed prior are very close to the
posterior probabilities had the analyst specified a prior comprising all
possibilities from the outset. It could be objected, though, that also even
further models could have been considered (that could have been chosen in case 
the original model would have been rejected for different reasons), and 
as this was in fact not done,
it cannot be known what implications this would have had. This can be taken 
as a reason to not change the prior based on the data too lightly.

The topic here is the use of epistemic probabilities in statistical data 
analysis. Philosophers interested in a general formal logic of plausibility
will use epistemic probabilities somewhat differently, even if they follow the
same Bayesian formalism with reference to Cox's Theorem. \cite{Jaynes03} 
motivates his introduction to Bayesian probability theory with equipping a 
robot with the formal tools to do automated reasoning. A robot cannot
be expected to take axioms and their implications ``with a large pinch of 
salt'', as a human statistician can, and such an attitude is probably not to
most philosophers' tastes either. A robot can be taught handling of epistemic
probabilities, but not (at least not currently) 
epistemic-probabilities-as-model, flexibly distancing itself from the 
absoluteness of a formal calculus and violating (or re-adjusting) it for a 
variety of good reasons that may not all be obvious in advance.    
  
In \cite{GelHen17} we have argued that it is worthwhile in science to aim
at aspects of objectivity such as transparency, consensus, and impartiality. 
Following a consistent formal system is surely attractive in these 
respects, but there is always an irreducible tension between these aims and the 
context-dependence of reasoning and the recognition of multiple perspectives.
We should be prepared to question and at least temporarily suspend any 
consistent formal system if required. acknowledging it as a model, an 
idealisation with limitations. A major issue with 
\hbindex{artificial intelligence} (AI)
(like Jaynes's robot) is that AI has to rely on formal models and cannot 
question them, or only in the framework of a bigger formal model.  

\subsection{How different are epistemic-probability-as-model and frequentism-as-model?}

In frequentism-as-model, an agent adopts a probability model temporarily, 
acknowledging its essential difference from reality, which
means that she considers the modelled experiment as if it has a tendency to 
give rise to the corresponding relative frequencies under repetition. In
epistemic-probability-as-model, an agent adopts a probability 
model temporarily, 
acknowledging its essential difference from reality, which
means that she considers the uncertainties in the modelled experiment as if,
were she to bet on future observations, she would do this according to
the model. 

Involving exchangeability (which of course is questionable) in the epistemic
model actually implies that relative frequencies are expected to converge, and
using a frequentist model could then be justified just as well, although 
a Bayesian may want to keep the \hbindex{prior}, 
at least if it encodes relevant 
information that would otherwise be lost. It has also been recommended in 
various
places \citep[Ch. 1]{GCSDVR13} to think about a situation as repeatable, 
and to consider expected relative frequencies when setting up Bayesian priors. 
In fact, Bayesians often think of the sampling model part of the prior as
modelling an aleatory mechanism (possibly idealised, in an ``as-model'' manner),
and often posterior probabilities regarding a
parameter are interpreted as if this were about a ``true'' parameter value 
defined within an aleatory sampling model. 

Taking also into account that epistemic and aleatory probabilities were 
historically not distinguished, frequentism-as-model
and epistemic-probability-as-model are indeed strongly related. It will however
contribute to the clarity of communication to specify, when doing analyses,
explicitly what the involved probabilities are supposed to mean. Does the
modeller model a \hbindex{data generating mechanism} in the world, and
do probabilities refer to the outcomes of it? Or do probabilities refer to
the agent's \hbindex{uncertainty}? 
I have elaborated that this is not an issue of 
belief in the existence of objective aleatory probabilities, which is not
more or less problematic than belief in a general epistemic calculus formalising
rational handling of uncertainty, and in specific prior choices, even if the 
latter only refer to the agent herself. Modelling does not require belief 
in the truth of a model; a model has to be justified from its use.
   
Aleatory models can in principle, by bold idealisation, even be applied to 
situations that are essentially non-repeatable and may not even allow for 
repeatable residual or innovation terms and the like. However, anchoring
such applications in observations and defending them convincingly may be 
hard if not impossible, and a modeller has an easier task for sure 
justifying the use of epistemic probabilities. 

On the other hand, many sequences of experiments are treated in probability 
modelling as if they behave like aleatory processes, and results are 
interpreted in this way, even if used as \hbindex{sampling model} 
in a Bayesian prior. It seems to be more direct to me in such cases
to talk about models for data generating processes in reality, 
rather than
about models for how an agent thinks about  data generating processes in 
reality,
and therefore I often prefer the aleatory interpretation, not ruling out
using Bayesian calculus. 

\section{Conclusion} \label{sconc}
It may seem to be my core message that models are models and as 
such different from reality. This is of course commonplace, and agreed by
many if not all statisticians, although it rarely influences applied statistical
analyses or even discussions about the foundations of statistics.

Here are some less obvious implications: 
\begin{itemize}
\item The usual way of talking about \hbindex{model assumptions}, namely 
that they ``have
to be fulfilled'', is misleading. The aim of model assumption checking is not to
make sure that they are fulfilled, but rather to rule out issues that misguide
the interpretation of the results. Combining model assumption checking and 
analyses chosen conditionally on the model checking results can itself be
modelled and analysed, and depending on what exactly is done, it may or
may not turn out to work well.  
\item There are always lots of models \hbindex{compatible} 
with the data. Some of these
can be favoured or excluded by plausibility considerations, prior information, 
or by the data, but some irregular ones have to 
be excluded simply because inference would be hopeless if they were correct. 
\item \hbindex{Model assessment} 
based on the data involves decisions about ``in what 
way to look'', i.e., what model deviations are relevant.  
Whether a model is compatible with the data cannot be decided 
independently of such considerations.
\item Choosing a model implies decisions to ignore certain aspects of reality,
e.g., differences between the conditions under which observations modelled as
i.i.d. were gathered. These decisions should be transparent and open for 
discussion. 
\item Consideration of \hbindex{interpretative equivalence}, i.e., what 
different models
would be interpreted in the same or different way regarding the subject matter,
are important in order to investigate \hbindex{robustness} and \hbindex{stability}, i.e., to what
extent different models compatible with the data 
would lead to results on the same data that
have a different meaning.
\item Frequentism-as-model is compatible with both compatibility and \hbindex{inverse
probability logic}. A key to decide which one to prefer is to ask whether the 
\hbindex{parameter prior} distribution required for inverse
probability logic can be used to add valuable information in a 
convincing way. The parameter 
prior itself can not normally be 
checked against the data with satisfactory power. Compatibility logic can also
be used for epistemic-probability-as-model; a model may be dropped that turns 
out to be incompatible with the observations, even thought this violates 
coherence.
\end{itemize}
{\bf Acknowledgements:} I thank Laurie Davies, Andrew Gelman, Donald Gillies, Sander Greenland, David Hand, Johannes H\"{u}sing, and Keith O'Rourke for helpful comments on an earlier version of this chapter.  

\bibliographystyle{chicago}

\end{document}